\documentclass[aps,prl,twocolumn,superscriptaddress,showpacs]{revtex4-1}

\usepackage{graphicx}
\usepackage{amsmath}

\newcommand{\eqn}[1]{
\begin{eqnarray}
	#1
\end{eqnarray}
}

\begin{document}


\title{Diffraction-Unlimited Position Measurement of Ultracold Atoms in an Optical Lattice}


\author{Yuto Ashida}
\affiliation{Department of Physics, University of Tokyo, 7-3-1 Hongo, Bunkyo-ku, Tokyo
113-0033, Japan}
\author{Masahito Ueda}
\affiliation{Department of Physics, University of Tokyo, 7-3-1 Hongo, Bunkyo-ku, Tokyo
113-0033, Japan}
\affiliation{RIKEN Center for Emergent Matter Science (CEMS), Wako, Saitama 351-0198, Japan
}

\date{\today}

\begin{abstract} 
We consider a method of high-fidelity, spatially resolved position measurement of ultracold atoms in an optical lattice. We show that the atom-number distribution can be nondestructively determined at a spatial resolution beyond the diffraction limit by tracking the progressive evolution of the many-body wavefunction collapse into a Fock state. We predict that the Pauli exclusion principle accelerates the rate of wavefunction collapse of fermions in comparison with bosons. A possible application of our principle of surpassing the diffraction limit to other imaging systems is discussed.
\end{abstract}


\maketitle

The classical theory of electromagnetism predicts that two objects with a distance less than a wavelength cannot be resolved \cite{BM99}. This fundamental limit known as the diffraction limit has long imposed insurmountable constraints on optical physics. Recent  achievements of the single-site resolved imaging \cite{BWS09,SJF10} and addressing \cite{WC11} of ultracold lattice gases are not exceptions: the diffraction limit requires a high numerical aperture lens and a large number of signals which forces us to use a near-resonant probe light which causes destruction of atomic states. As a result, all the experiments on single-site-resolved detection performed to date are destructive. 

Quantum gases in an optical lattice offer an ideal playground to investigate strongly correlated systems and quantum information \cite{JDB98,ML07,IB08}. Recently, the single-site resolved detection and addressing have emerged as a powerful tool for those studies \cite{BWS10,SJ11,EM11,MC12,FT13,FT132,PMP15,FT15}. Against such a backdrop, the development of a nondestructive measurement at the single-site level will have a significant impact on the field of quantum simulation \cite{BI12,GIM14}, quantum information processing \cite{NKD07,BT12}, and open quantum many-body systems \cite{DAJ14,PYS14}. Furthermore, it will also be applied to the study of influence of measurement back-action on quantum many-body states \cite{SK12,DAJ12,DJS12,SK13,PYS142} and open up the possibility of generalizing the concepts in quantum feedback control \cite{WHM10,SC11} and quantum non-demolition measurement (QND) \cite{BM90,GC07,RT14} to quantum many-body systems. To move toward these goals, there is an obvious need to develop methods which allow us to overcome the difficulty posed by the diffraction limit.

In this Letter, we propose a method that achieves this aim. We consider a quantum measurement of atomic positions in an optical lattice by spatially-resolved detections of dispersively scattered photons. A photodetection induces the many-body wavefunction collapse of atoms because a lens aperture diffracts the scattered field and imprints the spatial information of atoms on photons \cite{HM96}.
We show that this measurement back-action will localize the atom-number distribution and that tracking the progressive collapse into a Fock state enables us to perform diffraction-unlimited position measurement with near-unit fidelity. 
The main idea of our scheme of surpassing the diffraction limit is as follows. Since quantum measurement theory automatically takes into account the method of Bayesian inference, we can extract the unbiased positional information of atoms and, remarkably, can precisely distinguish atomic configurations even if the diffraction limit does not reach the lattice constant. Previous works discussing nondestructive methods using off-resonant scattering with the angle-resolved measurement \cite{JJ95,JJ952,IZ00,RAM05,LK09,DJS11} and with the use of a cavity \cite{MIB07,MIB09} cannot achieve such a high spatial resolution. 
Furthermore, we find that the Pauli exclusion principle accelerates the rate of wavefunction collapse of fermions compared with bosons and, thus, our scheme is particularly advantageous for the recently realized single-site detection of fermionic gases \cite{CLW15,PMF15,EH15}.

While we here focus on atoms, our scheme can also be applied to other optical lattice systems such as single trapped ions \cite{LD03,SEW11,AJ11}. Further, since the Bayesian analysis of the conditional probability distribution of source positions allows us to surpass the diffraction limit, our formulation may have an application to other imaging systems, e.g., photoactivated  localization microscopy \cite{MWE89,DRM97,BE06,RMJ06,SJS13,YA15} as discussed later. We note that there also exist other approaches of overcoming the diffraction limit such as sub-wavelength-scale optical lattices using plasmons \cite{GM12}, photonic crystals \cite{AGT15}, and running waves \cite{SN15}. Other examples include stimulated emission depletion microscopy \cite{KTA00} and its coherent extension to magnetic imaging \cite{DLS13}.



{\it Model.---}
We consider two-level atoms in a lattice described by the many-body Hamiltonian 

\begin{figure}[t]
\includegraphics[scale=0.36]{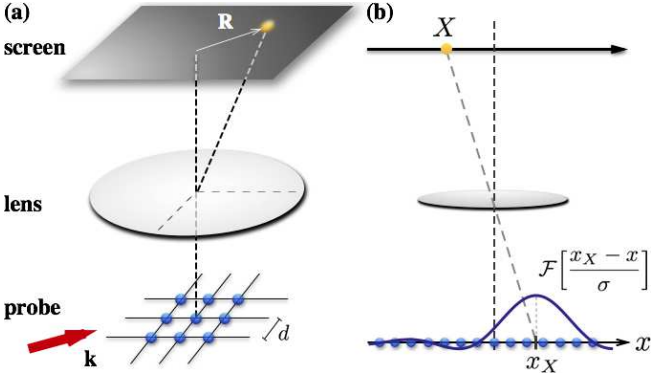}
\caption{\label{fig1} 
(color online). Schematic geometry of our system. (a) Atoms trapped in an optical lattice are illuminated by an off-resonant probe light with wave vector $\mathbf{k}$. A scattered light is diffracted through a lens and detected on the screen. The position of a detected photon is denoted by $\mathbf{R}$. (b) A measurement back-action caused by the detection of a photon at position $X$. The many-body wavefunction shrinks according to the function $\mathcal{F}$ which peaks at $x_{X}$, where $x_{X}$ is the lattice point diametrically opposite to $X$ with respect to the center of the lens aperture.
}
\end{figure}

\eqn{
\hat{\mathcal{H}}=\int d^{3}r\Bigl[\hat{H}_{a}(\mathbf{r})+\hat{H}_{af}(\mathbf{r})\Bigr]+\hat{H}_{f},
}
where $\hat{H}_{a}(\mathbf{r})=\sum_{i=g,e}\hbar\omega_{i}\hat{\Psi}_{i}^{\dagger}(\mathbf{r})\hat{\Psi}_{i}(\mathbf{r})$ is the Hamiltonian of atoms and $ \hbar\omega_{g,e}$ are their ground ($g$) and excited ($e$) state energies, and $\hat{\Psi}_{g,e}(\mathbf{r})$ are the corresponding field operators; $\hat{H}_{af}(\mathbf{r})=-(\mathbf{d}\cdot\hat{\mathbf E}(\mathbf{r})\hat{\Psi}_{g}^{\dagger}(\mathbf{r})\hat{\Psi}_{e}(\mathbf{r})+\rm{H.\,c.\,})$ describes the electric-dipole interaction with $\mathbf{d}$ and $\hat{\mathbf{E}}(\mathbf{r})$ being the electric dipole moment and the electric field operator, respectively; $\hat{H}_{f}=\sum_{\mathbf{k'},\sigma}\hbar\omega_{k'}\hat{a}^{\dagger}_{\mathbf{k'},\sigma}\hat{a}_{\mathbf{k'},\sigma}$ is the free-field Hamiltonian with $\hat{a}_{\mathbf{k'},\sigma}$ being the annihilation operator of a photon with wave vector $\mathbf{k'}$ and polarization $\sigma$. Atoms are illuminated by an off-resonant probe light whose positive frequency component is $\mathbf{E}_{\rm P}^{(+)}(\mathbf{r})=\mathbf{e}_{\rm P}\mathcal{E}_{0}e^{i\mathbf{k}\cdot\mathbf{r}}/2$. 
Each scattered photon is diffracted through a lens aperture and detected on a screen (see Fig. \ref{fig1}(a)). We first focus on a 1D lattice and then discuss the generalization to a 2D lattice. In this Letter, we ignore tunneling of atoms through lattice potentials during imaging and focus on light scattering. 

The scattered field can be calculated \cite{SM1} by integrating out the Heisenberg equation of motion under the geometry as shown in Fig. \ref{fig1}.  After performing the adiabatic elimination of the excited state and employing the tight-binding approximation, we arrive at the following expression of the scattered field at position $X$ on the screen:
 \eqn{\label{Sca1D}
 \hat{E}_{\rm sca}^{(+)}(X)=\gamma\sum_{m}e^{-i\Delta\mathbf{k}\cdot md\mathbf{e}_{x}}\mathcal{F}\Bigl[\frac{|x_{X}-md|}{\sigma}\Bigr]\hat{b}^{\dagger}_{m}\hat{b}_{m},
 }
where the polarization vector is averaged out since we do not consider measuring the polarization of a photon, $\Delta\mathbf{k}$ is the wave-vector difference between incident and scattered photons, the operator $\hat{b}_{m}$ annihilates an atom at site $m$, $x_{X}$ is the diagonal coordinate of the detected position (Fig. \ref{fig1}(b)), and $\mathcal{F}[y]\equiv J_{1}(y)/y$ which vanishes rapidly for $y\gg1$. Here we assume that atoms remain in the lowest-band Wannier states during the imaging. This assumption should be met by using, for example, Raman sideband cooling \cite{PYS14,PYS142,CLW15,PMF15} as discussed later. We introduce the parameter $\sigma$ characterizing the resolution of the classical imaging method which is defined by the numerical aperture of the lens $N_{\rm A}$ as $\sigma\equiv1/kN_{\rm A}$. The diffraction limit is usually characterized by the first zero of the Airy disk, $d_{\rm diff}=0.61\frac{\lambda}{N_{\rm A}}$, which can be related to $\sigma$ as $d_{\rm diff}=3.8 \sigma$. Note that the  classical imaging method can achieve the single-site resolved measurement only when the diffraction limit reaches the scale of the lattice constant $d_{\rm diff}\lesssim d$.

Let us now consider the physical content of Eq. (\ref{Sca1D}). The measured observable can be continuously varied as we control the parameter $\sigma$ by, for example, changing the distance between a lens and a lattice. When the lens is positioned at a far-field region and the numerical aperture is so low that $\sigma$ is much larger than the lattice constant, the phase factor in Eq. (\ref{Sca1D}) generates a Bragg diffraction pattern rather than the space-resolved imaging \cite{WC112}. 

We develop a continuous quantum measurement theory and show that successive photodetecions cause progressive collapse of the atomic state into a Fock state which, in turn, allows us to surpass the diffraction limit. 
We first consider an ideal situation in which the collection efficiency of scattered photons is unity and later discuss the effect of uncollected photons. Since the effective Hamiltonian commutes with the atom-number operator at each lattice site, the ideal photodetection of a dispersive scattered light constitutes a QND measurement of the atom-number statistics in the sense of Ref. \cite{BVB96}. Hence, our model presents a dual approach compared with a QND model \cite{BM90,GC07,RT14} of the photon number based on two-level atoms, where the photon number is determined by detecting the state of an output atom.

For the sake of concreteness, let us consider $N$ atoms trapped in a 1D optical lattice with $N_{\rm L}$ sites. The state of a quantum gas is represented in terms of Fock states $|\{n_{m}\}\rangle\equiv|n_{1},\ldots,n_{N_{\rm L}}\rangle$ satisfying $\sum_{m=1}^{N_{\rm L}}n_{m}=N$.  Let $\rho_{0}$ be the density matrix for the initial motional state of atoms and $P_{0}[\{n_{m}\}]$ be the corresponding initial atom-number distribution. When we detect a photon at the screen position $X$, the change of the conditional state can be described by the measurement operator (\ref{Sca1D}) as $\rho_{0}\to\hat{E}^{(+)}_{\rm sca}(X)\rho_{0}\hat{E}^{\dagger(+)}_{\rm sca}(X)/{\rm Tr}[\hat{E}^{(+)}_{\rm sca}(X)\rho_{0}\hat{E}^{\dagger(+)}_{\rm sca}(X)]$.

Suppose now that $n$ photons were detected at the positions $\mathbf{X}\equiv\{X_{1},\ldots,X_{n}\}$. Then the atom-number distribution of the quantum state becomes
\eqn{\label{SC}
P_{n}[\{n_{m}\}|\mathbf{X}]=\frac{P_{0}[\{n_{m}\}]\prod_{k=1}^{n}P[X_{k}|\{n_{m}\}]}{\sum_{\{n'_{m}\}}P_{0}[\{n'_{m}\}]\prod_{k=1}^{n}P[X_{k}|\{n'_{m}\}]}.\nonumber\\
}
Here $P[X|\{n_{m}\}]$ is the conditional probability of detecting a photon at $X$, given that the atomic state is the Fock state $|\{n_{m}\}\rangle$:
\eqn{\label{Cp}
P[X|\{n_{m}\}]=\frac{\bigl|\sum_{m=1}^{N_{\rm L}}n_{m}\mathcal{F}\bigl[\frac{|x_{X}-md|}{\sigma}\bigr]\bigr|^{2}}{\int dX'\bigl|\sum_{m=1}^{N_{\rm L}}n_{m}\mathcal{F}\bigl[\frac{|x_{X'}-md|}{\sigma}\bigr]\bigr|^{2}},
}
where we neglect the contribution of the phase factor \cite{SM1} in Eq. (\ref{Sca1D}). We can track the progressive dynamics of the wavefunction collapse by applying Eq. (\ref{SC}) iteratively. The detection of a sufficiently large number of photons causes the state to collapse into a Fock state $|\{n_{m}\}\rangle$ and, hence, the occupied atom number at each lattice will be precisely determined. Note that the order of the measurement outcomes is irrelevant to the final collapsed state because all measurement operators commute with each other and only the accumulated histogram of the positional information of photodetections is sufficient to determine the final atom-number distribution. Nevertheless, tracking the progressive collapse in real time allows an adaptive measurement to terminate an imaging process once the required confidence level is attained so that unfavorable effects such as heating are kept minimal.

Let us here discuss why our method can surpass the diffraction limit. The crucial point is that since Eq. (\ref{SC}) derived from the quantum measurement theory automatically takes into account the  Bayesian inference, we can extract the detailed unbiased information about the positions of point sources. While the atomic configuration can also be inferred to some extent from an ordinary fitting procedure with a sufficiently large number of photodetections, our scheme achieves much better accuracy as detailed later. 

\begin{figure}
\scalebox{0.24}[0.24]{\includegraphics{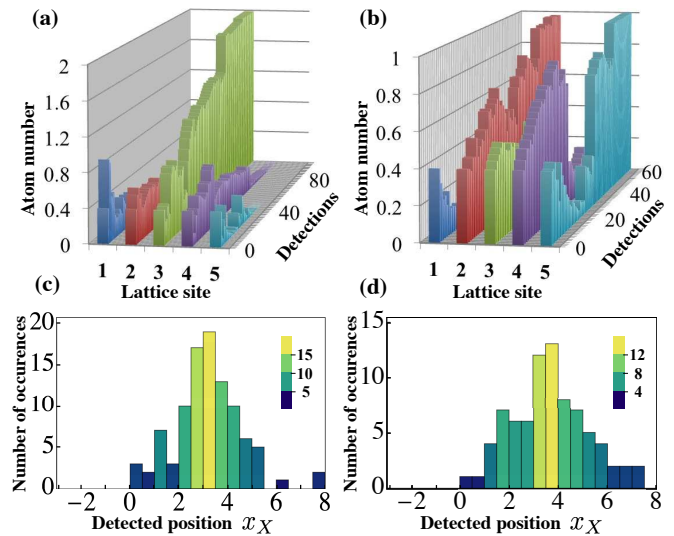}}
 \caption{\label{fig2}
(color online). Wavefunction collapse of atoms into Fock states (a,b) and the diffraction-limited imaging signals (c,d). (a,b) Successive detections of photons induce the state reduction of (a) bosons and (b) fermions into Fock states, with $N_{\rm L}=5, N=2, \sigma=1.0$ (in units of $d$). The atom number at each site is determined with near-unit fidelity by tracking the wavefunction collapse. (c,d) The associated histograms of photodetection positions for (c) bosons and (d) fermions, respectively. Signals from atoms placed at different sites are masked by diffractions.
}
\end{figure}

{\it Numerical simulations.---}
To illustrate the principle of surpassing the diffraction limit, we perform numerical simulations of our model. Figures \ref{fig2}(a) and (b) respectively show the collapse of a bosonic state and that of a fermonic state into Fock states, where the initial state is chosen as a superposition of all possible Fock states. For comparison, we also show the histograms of detected positions of photons (Figs. \ref{fig2}(c) and (d)). Our method enables us to distinguish between different Fock states by tracking the wavefunction collapse and hence, we can determine the atom number at each lattice site with near-unit fidelity beyond the conventional parity measurement. (For possible collision-induced loss, see the discussion below.) On the other hand, the classical diffraction-limited images (Figs. \ref{fig2}(c) and (d)) cannot resolve atoms placed at neighboring sites.  The distribution of the collapsed states obtained by many realizations reproduces the initial state distribution.

\begin{figure}[t]
\scalebox{0.25}[0.25]{\includegraphics{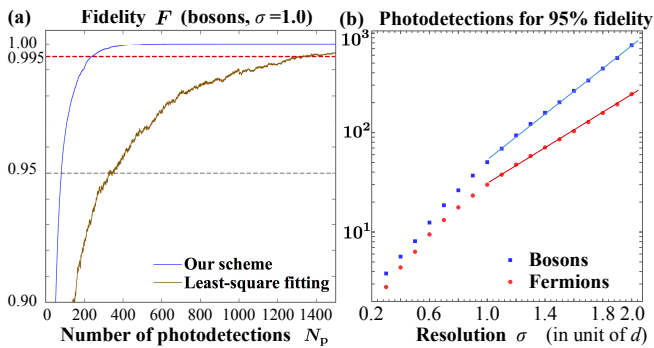}} 
\caption{\label{fig3}
(color online). Fidelity $F$ and the required number of photodetections $N_{\rm p}$. (a) The fidelity of our scheme (blue) compared with the result of the least-square fitting (yellow) for bosons at $\sigma=1.0$. Dashed lines indicate 95\% and 99.5\% confidence level. (b)  The required number of photodetections to achieve 95\% fidelity (log-scale) plotted against the resolution $\sigma$ for bosons (blue) and fermions (red), calculated for $10^{4}$ realizations with $N_{\rm L}=5$ and $N=2$.
}
\end{figure}

To show the high fidelity of our scheme, we plot the confidence level of identifying the collapsed Fock state against the number of photodetections (Fig. \ref{fig3}(a)). As shown, our scheme achieves a faster convergence to unit-fidelity than an ordinary  fitting procedure. In particular, we find about an order of magnitude improvement in the number of photodetections at 99.5\% fidelity, which is the confidence level reported in Ref. \cite{SJF10}. We note that the regular analysis here also utilizes the same knowledge about the discrete positions in the lattice \cite{SM2}.

To investigate how quantum statistics of atoms affects the evolution of the wavefunction collapse, we plot the rate of wavefunction collapse against the resolution parameter $\sigma$ (Fig. \ref{fig3}(b)).  We find that the rate of convergence is faster for fermions than bosons. This can be attributed to the Pauli exclusion principle which greatly reduces the number of possible configurations for fermions. Hence, our method is particularly advantageous for the single-site resolved detections of fermionic gases \cite{CLW15,PMF15,EH15}. 
Another interesting feature is that the average number of photodetections needed to cause the many-body wavefunction collapse grows almost exponentially with the resolution parameter $\sigma$ in a large-$\sigma$ region. Because the rate of convergence can be related to the relative entropy of each measurement \cite{BM11}, this finding should have some information-theoretic background. 

{\it Asymptotic formula of the fidelity.---}
To explicitly show the reach of our scheme, here we discuss an analytical treatment of the confidence level. The fidelity $F$ of identifying the collapsed Fock state can be asymptotically approximated by the following equation \cite{BM11}:
\eqn{\label{FD}
F=1-\sum_{\{n_{m}\}}P_{0}[\{n_{m}\}]e^{-N_{\rm p}R[\{n_{m}\}]},
}
where $N_{\rm p}$ is the total number of photodetections and $R[\{n_{m}\}]$ is the minimal information-theoretic distance of the Fock state $\{n_{m}\}$:
\eqn{\label{KL}
R[\{n_{m}\}]\equiv\min_{\{n_{k}\}\neq\{n_{m}\}}D[P[X|\{n_{m}\}]|P[X|\{n_{k}\}]].
}
Here $D[P_{1}|P_{2}]$ denotes the relative entropy between $P_{1,2}$.  Equation (\ref{FD}) has the following physical meaning: the more distinguishable the interference patterns generated by different Fock states are, the faster the convergence to unit-fidelity occurs. From Eq. (\ref{FD}), one can calculate the expected fidelity for an arbitrary number of atoms and lattice sites. A scaling analysis of the error-rate $\epsilon\equiv1-F$ implies that the fidelity scales mainly with the number of detections per atom (per lattice site) for fermions (bosons) \cite{SM2}. 

As inferred from Eq. (\ref{FD}), an order-of-magnitude improvement of $N_{\rm p}$ (Fig. \ref{fig3}(a)) allows exponential improvement of the precision i.e., the error-rate of finding a particular Fock state becomes exponentially small. This is particularly advantageous feature of the discretized space compared with the continuous space in which case only the square-root improvement of the precision is expected. 

{\it Experimental situations.---}
Let us here make some practical considerations on experimental implementation. First, we note that, in practice, only a portion of scattered photons can be collected and the back-action caused by destructions of uncollected photons in the far-field region also affects the dynamics of the wavefunction collapse \cite{DJS12,RAV03}. In this case, Eq. (\ref{SC}) does not reconstruct an exact progressive dynamics of the wavefunction collapse. Nevertheless, the actual final collapsed Fock state, which is of primary interest, coincides with the one identified from Eq. (\ref{SC}) after a sufficient number of photodetections. As we pointed out earlier, the order of photodetection does not matter for the eventual wavefunction collapse; this implies that the presence of uncollected photons only delays the speed of the collapse but not alter the eventual atomic state. Hence, our method can determine the atom distribution even in the presence of uncollected photons. In this respect, our formulation based on the Bayesian update of the conditional probability distribution of source positions may have an application to super-resolved imaging of classical objects such as photoactivated  localization microscopy \cite{MWE89,DRM97,BE06,RMJ06,SJS13,YA15}.

Second, we note that our scheme can, in principle, be performed without prior knowledge of the total atom number \cite{KM09,WAN13}. This is because the relative entropy $D[P_{\rm meas}[X]|P[X|\{n_{m}\}]]$ between the measured photon-number distribution and the distribution from the collapsed Fock state provides a way of hypothesis testing of the atom number \cite{SM2}.

Third, we estimate a possible heating and show appropriate experimental parameters for our scheme. To this end, we consider the setup of Ref. \cite{SJF10}: $d=532\rm{nm}$, $\lambda=780\rm{nm}$ and $N_{A}=0.68$ (leading to $\sigma=0.343$). Combining our numerical results of the required number of detections (about 5 detections per atom at $\sigma=0.343$) with the experimental collection efficiency $\sim$10\%, we obtain total recoil energy as $\sim93E_{r}$ where $E_{r}\equiv \hbar^{2}\pi^{2}/2md^{2}$. Hence, the contribution of heating effects would be made negligible by, for example, implementing Raman sideband cooling as recently performed in Refs. \cite{CLW15,PMF15}. To estimate the imaging time $T_{\rm img}$, we consider the $\rm D_{2}$-transition of $^{87}\rm Rb$ \cite{SJF10} and the detuning $\Delta=100\Gamma$, where $\Gamma/2\pi=6.07 {\rm MHz}$ is the decay rate. From the required number of photodetections combined with the collection efficiency mentioned above, we obtain $T_{\rm img}\sim9.6 {\rm ms}$.

As for a possible loss of bosons due to light-assisted inelastic collisions, we note that usage of an off-resonant blue-detuned light can suppress such radiative losses \cite{JW03}. To quantify the argument, we evaluate the inelastic scattering rate based on the Landau-Zener transition probability \cite{JW03}. From the above experimental parameters combined with $\omega_{\rm trap}/2\pi=20{\rm kHz}$ for the harmonic trapping frequency of the optical lattice, we obtain $6\times 10^{-3} {\rm Hz}$ as the inelastic collision rate per atom and hence, radiative losses can indeed be neglected in our consideration. We note that the Doppler heating effect can also be neglected because we consider large detuning. The crucial point is that, our method allows us to distinguish different Fock states with much less photons and thus without substantial heating which, in turn, allows the usage of a blue-detuned light. This contrasts with the conventional methods \cite{BWS09,SJF10}, which require use of a near-resonant red-detuned light to simultaneously achieve high scattering rate and cooling atoms. 

Finally, while reconstructing the collapse dynamics of a mesoscale number of atoms (typical system in quantum gas microscope experiments  \cite{MC12,FT13,FT132,PMP15,FT15}) are already within the scope of our model, one may need to reconstruct a macroscopic number of atoms, which, in practice, seems to be beyond the scope due to an exponential growth of computational cost. We note, however, that if the final collapsed Fock state is of primary interest, one can avoid such problems by neglecting an intermediate process of the collapse \cite{YA15}.      

{\it Conclusion.---}
We have demonstrated that tracking the progressive evolution of wavefunction collapse of a quantum many-body state provides a way to surpass the classical resolution limit, which---in contrast to the conventional diffraction-limited parity measurement---enables a nondestructive measurement of the atom-number distribution at the single-site level. Moreover, our principle of surpassing the diffraction limit has a much broader range of applications other than an optical lattice system and gives a powerful means to extract positional information in different varieties of challenging situations.

We acknowledge I. Bloch, T. Fukuhara, R. Inoue, M. Miranda and S. Nakajima for valuable discussions. This work was supported by
KAKENHI Grant No. 26287088 from the Japan Society for the Promotion of Science, 
and a Grant-in-Aid for Scientific Research on Innovative Areas ``Topological Materials Science" (KAKENHI Grant No. 15H05855),
and the Photon Frontier Network Program from MEXT of Japan, and the Mitsubishi Foundation. Y. A. was supported by the Japan Society for the Promotion of Science through Program for Leading Graduate Schools (ALPS). 

\bibliography{reference}

\begin{thebibliography}{66}%
\makeatletter
\providecommand \@ifxundefined [1]{%
 \@ifx{#1\undefined}
}%
\providecommand \@ifnum [1]{%
 \ifnum #1\expandafter \@firstoftwo
 \else \expandafter \@secondoftwo
 \fi
}%
\providecommand \@ifx [1]{%
 \ifx #1\expandafter \@firstoftwo
 \else \expandafter \@secondoftwo
 \fi
}%
\providecommand \natexlab [1]{#1}%
\providecommand \enquote  [1]{``#1''}%
\providecommand \bibnamefont  [1]{#1}%
\providecommand \bibfnamefont [1]{#1}%
\providecommand \citenamefont [1]{#1}%
\providecommand \href@noop [0]{\@secondoftwo}%
\providecommand \href [0]{\begingroup \@sanitize@url \@href}%
\providecommand \@href[1]{\@@startlink{#1}\@@href}%
\providecommand \@@href[1]{\endgroup#1\@@endlink}%
\providecommand \@sanitize@url [0]{\catcode `\\12\catcode `\$12\catcode
  `\&12\catcode `\#12\catcode `\^12\catcode `\_12\catcode `\%12\relax}%
\providecommand \@@startlink[1]{}%
\providecommand \@@endlink[0]{}%
\providecommand \url  [0]{\begingroup\@sanitize@url \@url }%
\providecommand \@url [1]{\endgroup\@href {#1}{\urlprefix }}%
\providecommand \urlprefix  [0]{URL }%
\providecommand \Eprint [0]{\href }%
\providecommand \doibase [0]{http://dx.doi.org/}%
\providecommand \selectlanguage [0]{\@gobble}%
\providecommand \bibinfo  [0]{\@secondoftwo}%
\providecommand \bibfield  [0]{\@secondoftwo}%
\providecommand \translation [1]{[#1]}%
\providecommand \BibitemOpen [0]{}%
\providecommand \bibitemStop [0]{}%
\providecommand \bibitemNoStop [0]{.\EOS\space}%
\providecommand \EOS [0]{\spacefactor3000\relax}%
\providecommand \BibitemShut  [1]{\csname bibitem#1\endcsname}%
\let\auto@bib@innerbib\@empty
\bibitem [{\citenamefont {Born}\ and\ \citenamefont {Wolf}(1999)}]{BM99}%
  \BibitemOpen
  \bibfield  {author} {\bibinfo {author} {\bibfnamefont {M.}~\bibnamefont
  {Born}}\ and\ \bibinfo {author} {\bibfnamefont {E.}~\bibnamefont {Wolf}},\
  }\href@noop {} {\emph {\bibinfo {title} {Principles of Optics}}}\ (\bibinfo
  {publisher} {Cambridge University Press},\ \bibinfo {address} {Cambridge,
  England},\ \bibinfo {year} {1999})\BibitemShut {NoStop}%
\bibitem [{\citenamefont {Bakr}\ \emph {et~al.}(2009)\citenamefont {Bakr},
  \citenamefont {Gillen}, \citenamefont {Peng}, \citenamefont {F{\"o}lling},\
  and\ \citenamefont {Greiner}}]{BWS09}%
  \BibitemOpen
  \bibfield  {author} {\bibinfo {author} {\bibfnamefont {W.~S.}\ \bibnamefont
  {Bakr}}, \bibinfo {author} {\bibfnamefont {J.~I.}\ \bibnamefont {Gillen}},
  \bibinfo {author} {\bibfnamefont {A.}~\bibnamefont {Peng}}, \bibinfo {author}
  {\bibfnamefont {S.}~\bibnamefont {F{\"o}lling}}, \ and\ \bibinfo {author}
  {\bibfnamefont {M.}~\bibnamefont {Greiner}},\ }\href@noop {} {\bibfield
  {journal} {\bibinfo  {journal} {Nature}\ }\textbf {\bibinfo {volume} {462}},\
  \bibinfo {pages} {74} (\bibinfo {year} {2009})}\BibitemShut {NoStop}%
\bibitem [{\citenamefont {Sherson}\ \emph {et~al.}(2010)\citenamefont
  {Sherson}, \citenamefont {Weitenberg}, \citenamefont {Endres}, \citenamefont
  {Cheneau}, \citenamefont {Bloch},\ and\ \citenamefont {Kuhr}}]{SJF10}%
  \BibitemOpen
  \bibfield  {author} {\bibinfo {author} {\bibfnamefont {J.~F.}\ \bibnamefont
  {Sherson}}, \bibinfo {author} {\bibfnamefont {C.}~\bibnamefont {Weitenberg}},
  \bibinfo {author} {\bibfnamefont {M.}~\bibnamefont {Endres}}, \bibinfo
  {author} {\bibfnamefont {M.}~\bibnamefont {Cheneau}}, \bibinfo {author}
  {\bibfnamefont {I.}~\bibnamefont {Bloch}}, \ and\ \bibinfo {author}
  {\bibfnamefont {S.}~\bibnamefont {Kuhr}},\ }\href@noop {} {\bibfield
  {journal} {\bibinfo  {journal} {Nature}\ }\textbf {\bibinfo {volume} {467}},\
  \bibinfo {pages} {68} (\bibinfo {year} {2010})}\BibitemShut {NoStop}%
\bibitem [{\citenamefont {Weitenberg}\ \emph
  {et~al.}(2011{\natexlab{a}})\citenamefont {Weitenberg}, \citenamefont
  {Endres}, \citenamefont {Sherson}, \citenamefont {Cheneau}, \citenamefont
  {Schau\ss}, \citenamefont {Fukuhara}, \citenamefont {Bloch},\ and\
  \citenamefont {Kuhr}}]{WC11}%
  \BibitemOpen
  \bibfield  {author} {\bibinfo {author} {\bibfnamefont {C.}~\bibnamefont
  {Weitenberg}}, \bibinfo {author} {\bibfnamefont {M.}~\bibnamefont {Endres}},
  \bibinfo {author} {\bibfnamefont {J.~F.}\ \bibnamefont {Sherson}}, \bibinfo
  {author} {\bibfnamefont {M.}~\bibnamefont {Cheneau}}, \bibinfo {author}
  {\bibfnamefont {P.}~\bibnamefont {Schau\ss}}, \bibinfo {author}
  {\bibfnamefont {T.}~\bibnamefont {Fukuhara}}, \bibinfo {author}
  {\bibfnamefont {I.}~\bibnamefont {Bloch}}, \ and\ \bibinfo {author}
  {\bibfnamefont {S.}~\bibnamefont {Kuhr}},\ }\href@noop {} {\bibfield
  {journal} {\bibinfo  {journal} {Nature}\ }\textbf {\bibinfo {volume} {471}},\
  \bibinfo {pages} {319} (\bibinfo {year} {2011}{\natexlab{a}})}\BibitemShut
  {NoStop}%
\bibitem [{\citenamefont {Jaksch}\ \emph {et~al.}(1998)\citenamefont {Jaksch},
  \citenamefont {Bruder}, \citenamefont {Cirac}, \citenamefont {Gardiner},\
  and\ \citenamefont {Zoller}}]{JDB98}%
  \BibitemOpen
  \bibfield  {author} {\bibinfo {author} {\bibfnamefont {D.}~\bibnamefont
  {Jaksch}}, \bibinfo {author} {\bibfnamefont {C.}~\bibnamefont {Bruder}},
  \bibinfo {author} {\bibfnamefont {J.~I.}\ \bibnamefont {Cirac}}, \bibinfo
  {author} {\bibfnamefont {C.~W.}\ \bibnamefont {Gardiner}}, \ and\ \bibinfo
  {author} {\bibfnamefont {P.}~\bibnamefont {Zoller}},\ }\href {\doibase
  10.1103/PhysRevLett.81.3108} {\bibfield  {journal} {\bibinfo  {journal}
  {Phys. Rev. Lett.}\ }\textbf {\bibinfo {volume} {81}},\ \bibinfo {pages}
  {3108} (\bibinfo {year} {1998})}\BibitemShut {NoStop}%
\bibitem [{\citenamefont {{Lewenstein}}\ \emph {et~al.}(2007)\citenamefont
  {{Lewenstein}}, \citenamefont {{Sanpera}}, \citenamefont {{Ahufinger}},
  \citenamefont {{Damski}}, \citenamefont {{Sen}},\ and\ \citenamefont
  {{Sen}}}]{ML07}%
  \BibitemOpen
  \bibfield  {author} {\bibinfo {author} {\bibfnamefont {M.}~\bibnamefont
  {{Lewenstein}}}, \bibinfo {author} {\bibfnamefont {A.}~\bibnamefont
  {{Sanpera}}}, \bibinfo {author} {\bibfnamefont {V.}~\bibnamefont
  {{Ahufinger}}}, \bibinfo {author} {\bibfnamefont {B.}~\bibnamefont
  {{Damski}}}, \bibinfo {author} {\bibfnamefont {A.}~\bibnamefont {{Sen}}}, \
  and\ \bibinfo {author} {\bibfnamefont {U.}~\bibnamefont {{Sen}}},\
  }\href@noop {} {\bibfield  {journal} {\bibinfo  {journal} {Adv. Phys.}\
  }\textbf {\bibinfo {volume} {56}},\ \bibinfo {pages} {243} (\bibinfo {year}
  {2007})}\BibitemShut {NoStop}%
\bibitem [{\citenamefont {Bloch}\ \emph {et~al.}(2008)\citenamefont {Bloch},
  \citenamefont {Dalibard},\ and\ \citenamefont {Zwerger}}]{IB08}%
  \BibitemOpen
  \bibfield  {author} {\bibinfo {author} {\bibfnamefont {I.}~\bibnamefont
  {Bloch}}, \bibinfo {author} {\bibfnamefont {J.}~\bibnamefont {Dalibard}}, \
  and\ \bibinfo {author} {\bibfnamefont {W.}~\bibnamefont {Zwerger}},\ }\href
  {\doibase 10.1103/RevModPhys.80.885} {\bibfield  {journal} {\bibinfo
  {journal} {Rev. Mod. Phys.}\ }\textbf {\bibinfo {volume} {80}},\ \bibinfo
  {pages} {885} (\bibinfo {year} {2008})}\BibitemShut {NoStop}%
\bibitem [{\citenamefont {Bakr}\ \emph {et~al.}(2010)\citenamefont {Bakr},
  \citenamefont {Peng}, \citenamefont {Tai}, \citenamefont {Ma}, \citenamefont
  {Simon}, \citenamefont {Gillen}, \citenamefont {Fölling}, \citenamefont
  {Pollet},\ and\ \citenamefont {Greiner}}]{BWS10}%
  \BibitemOpen
  \bibfield  {author} {\bibinfo {author} {\bibfnamefont {W.~S.}\ \bibnamefont
  {Bakr}}, \bibinfo {author} {\bibfnamefont {A.}~\bibnamefont {Peng}}, \bibinfo
  {author} {\bibfnamefont {M.~E.}\ \bibnamefont {Tai}}, \bibinfo {author}
  {\bibfnamefont {R.}~\bibnamefont {Ma}}, \bibinfo {author} {\bibfnamefont
  {J.}~\bibnamefont {Simon}}, \bibinfo {author} {\bibfnamefont {J.~I.}\
  \bibnamefont {Gillen}}, \bibinfo {author} {\bibfnamefont {S.}~\bibnamefont
  {Fölling}}, \bibinfo {author} {\bibfnamefont {L.}~\bibnamefont {Pollet}}, \
  and\ \bibinfo {author} {\bibfnamefont {M.}~\bibnamefont {Greiner}},\ }\href
  {\doibase 10.1126/science.1192368} {\bibfield  {journal} {\bibinfo  {journal}
  {Science}\ }\textbf {\bibinfo {volume} {329}},\ \bibinfo {pages} {547}
  (\bibinfo {year} {2010})}\BibitemShut {NoStop}%
\bibitem [{\citenamefont {Simon}\ \emph {et~al.}(2011)\citenamefont {Simon},
  \citenamefont {Bakr}, \citenamefont {Ma}, \citenamefont {Tai}, \citenamefont
  {Preiss},\ and\ \citenamefont {Greiner}}]{SJ11}%
  \BibitemOpen
  \bibfield  {author} {\bibinfo {author} {\bibfnamefont {J.}~\bibnamefont
  {Simon}}, \bibinfo {author} {\bibfnamefont {W.~S.}\ \bibnamefont {Bakr}},
  \bibinfo {author} {\bibfnamefont {R.}~\bibnamefont {Ma}}, \bibinfo {author}
  {\bibfnamefont {M.~E.}\ \bibnamefont {Tai}}, \bibinfo {author} {\bibfnamefont
  {P.~M.}\ \bibnamefont {Preiss}}, \ and\ \bibinfo {author} {\bibfnamefont
  {M.}~\bibnamefont {Greiner}},\ }\href@noop {} {\bibfield  {journal} {\bibinfo
   {journal} {Nature}\ }\textbf {\bibinfo {volume} {472}},\ \bibinfo {pages}
  {307} (\bibinfo {year} {2011})}\BibitemShut {NoStop}%
\bibitem [{\citenamefont {{Endres}}\ \emph {et~al.}(2011)\citenamefont
  {{Endres}}, \citenamefont {{Cheneau}}, \citenamefont {{Fukuhara}},
  \citenamefont {{Weitenberg}}, \citenamefont {{Schau{\ss}}}, \citenamefont
  {{Gross}}, \citenamefont {{Mazza}}, \citenamefont {{Ba{\~n}uls}},
  \citenamefont {{Pollet}}, \citenamefont {{Bloch}},\ and\ \citenamefont
  {{Kuhr}}}]{EM11}%
  \BibitemOpen
  \bibfield  {author} {\bibinfo {author} {\bibfnamefont {M.}~\bibnamefont
  {{Endres}}}, \bibinfo {author} {\bibfnamefont {M.}~\bibnamefont {{Cheneau}}},
  \bibinfo {author} {\bibfnamefont {T.}~\bibnamefont {{Fukuhara}}}, \bibinfo
  {author} {\bibfnamefont {C.}~\bibnamefont {{Weitenberg}}}, \bibinfo {author}
  {\bibfnamefont {P.}~\bibnamefont {{Schau{\ss}}}}, \bibinfo {author}
  {\bibfnamefont {C.}~\bibnamefont {{Gross}}}, \bibinfo {author} {\bibfnamefont
  {L.}~\bibnamefont {{Mazza}}}, \bibinfo {author} {\bibfnamefont {M.~C.}\
  \bibnamefont {{Ba{\~n}uls}}}, \bibinfo {author} {\bibfnamefont
  {L.}~\bibnamefont {{Pollet}}}, \bibinfo {author} {\bibfnamefont
  {I.}~\bibnamefont {{Bloch}}}, \ and\ \bibinfo {author} {\bibfnamefont
  {S.}~\bibnamefont {{Kuhr}}},\ }\href@noop {} {\bibfield  {journal} {\bibinfo
  {journal} {Science}\ }\textbf {\bibinfo {volume} {334}},\ \bibinfo {pages}
  {200} (\bibinfo {year} {2011})}\BibitemShut {NoStop}%
\bibitem [{\citenamefont {Cheneau}\ \emph {et~al.}(2012)\citenamefont
  {Cheneau}, \citenamefont {Barmettler}, \citenamefont {Poletti}, \citenamefont
  {Endres}, \citenamefont {Schau{\ss}}, \citenamefont {Fukuhara}, \citenamefont
  {Gross}, \citenamefont {Bloch}, \citenamefont {Kollath},\ and\ \citenamefont
  {Kuhr}}]{MC12}%
  \BibitemOpen
  \bibfield  {author} {\bibinfo {author} {\bibfnamefont {M.}~\bibnamefont
  {Cheneau}}, \bibinfo {author} {\bibfnamefont {P.}~\bibnamefont {Barmettler}},
  \bibinfo {author} {\bibfnamefont {D.}~\bibnamefont {Poletti}}, \bibinfo
  {author} {\bibfnamefont {M.}~\bibnamefont {Endres}}, \bibinfo {author}
  {\bibfnamefont {P.}~\bibnamefont {Schau{\ss}}}, \bibinfo {author}
  {\bibfnamefont {T.}~\bibnamefont {Fukuhara}}, \bibinfo {author}
  {\bibfnamefont {C.}~\bibnamefont {Gross}}, \bibinfo {author} {\bibfnamefont
  {I.}~\bibnamefont {Bloch}}, \bibinfo {author} {\bibfnamefont
  {C.}~\bibnamefont {Kollath}}, \ and\ \bibinfo {author} {\bibfnamefont
  {S.}~\bibnamefont {Kuhr}},\ }\href@noop {} {\bibfield  {journal} {\bibinfo
  {journal} {Nature}\ }\textbf {\bibinfo {volume} {481}},\ \bibinfo {pages}
  {484} (\bibinfo {year} {2012})}\BibitemShut {NoStop}%
\bibitem [{\citenamefont {{Fukuhara}}\ \emph {et~al.}(2013)\citenamefont
  {{Fukuhara}}, \citenamefont {{Kantian}}, \citenamefont {{Endres}},
  \citenamefont {{Cheneau}}, \citenamefont {{Schau{\ss}}}, \citenamefont
  {{Hild}}, \citenamefont {{Bellem}}, \citenamefont {{Schollw{\"o}ck}},
  \citenamefont {{Giamarchi}}, \citenamefont {{Gross}}, \citenamefont
  {{Bloch}},\ and\ \citenamefont {{Kuhr}}}]{FT13}%
  \BibitemOpen
  \bibfield  {author} {\bibinfo {author} {\bibfnamefont {T.}~\bibnamefont
  {{Fukuhara}}}, \bibinfo {author} {\bibfnamefont {A.}~\bibnamefont
  {{Kantian}}}, \bibinfo {author} {\bibfnamefont {M.}~\bibnamefont {{Endres}}},
  \bibinfo {author} {\bibfnamefont {M.}~\bibnamefont {{Cheneau}}}, \bibinfo
  {author} {\bibfnamefont {P.}~\bibnamefont {{Schau{\ss}}}}, \bibinfo {author}
  {\bibfnamefont {S.}~\bibnamefont {{Hild}}}, \bibinfo {author} {\bibfnamefont
  {D.}~\bibnamefont {{Bellem}}}, \bibinfo {author} {\bibfnamefont
  {U.}~\bibnamefont {{Schollw{\"o}ck}}}, \bibinfo {author} {\bibfnamefont
  {T.}~\bibnamefont {{Giamarchi}}}, \bibinfo {author} {\bibfnamefont
  {C.}~\bibnamefont {{Gross}}}, \bibinfo {author} {\bibfnamefont
  {I.}~\bibnamefont {{Bloch}}}, \ and\ \bibinfo {author} {\bibfnamefont
  {S.}~\bibnamefont {{Kuhr}}},\ }\href@noop {} {\bibfield  {journal} {\bibinfo
  {journal} {Nat. Phys.}\ }\textbf {\bibinfo {volume} {9}},\ \bibinfo {pages}
  {235} (\bibinfo {year} {2013})}\BibitemShut {NoStop}%
\bibitem [{\citenamefont {Fukuhara}\ \emph {et~al.}(2013)\citenamefont
  {Fukuhara}, \citenamefont {Schausz}, \citenamefont {Endres}, \citenamefont
  {Hild}, \citenamefont {Cheneau}, \citenamefont {Bloch},\ and\ \citenamefont
  {Gross}}]{FT132}%
  \BibitemOpen
  \bibfield  {author} {\bibinfo {author} {\bibfnamefont {T.}~\bibnamefont
  {Fukuhara}}, \bibinfo {author} {\bibfnamefont {P.}~\bibnamefont {Schausz}},
  \bibinfo {author} {\bibfnamefont {M.}~\bibnamefont {Endres}}, \bibinfo
  {author} {\bibfnamefont {S.}~\bibnamefont {Hild}}, \bibinfo {author}
  {\bibfnamefont {M.}~\bibnamefont {Cheneau}}, \bibinfo {author} {\bibfnamefont
  {I.}~\bibnamefont {Bloch}}, \ and\ \bibinfo {author} {\bibfnamefont
  {C.}~\bibnamefont {Gross}},\ }\href@noop {} {\bibfield  {journal} {\bibinfo
  {journal} {Nature}\ }\textbf {\bibinfo {volume} {502}},\ \bibinfo {pages}
  {76} (\bibinfo {year} {2013})}\BibitemShut {NoStop}%
\bibitem [{\citenamefont {Preiss}\ \emph {et~al.}(2015)\citenamefont {Preiss},
  \citenamefont {Ma}, \citenamefont {Tai}, \citenamefont {Lukin}, \citenamefont
  {Rispoli}, \citenamefont {Zupancic}, \citenamefont {Lahini}, \citenamefont
  {Islam},\ and\ \citenamefont {Greiner}}]{PMP15}%
  \BibitemOpen
  \bibfield  {author} {\bibinfo {author} {\bibfnamefont {P.~M.}\ \bibnamefont
  {Preiss}}, \bibinfo {author} {\bibfnamefont {R.}~\bibnamefont {Ma}}, \bibinfo
  {author} {\bibfnamefont {M.~E.}\ \bibnamefont {Tai}}, \bibinfo {author}
  {\bibfnamefont {A.}~\bibnamefont {Lukin}}, \bibinfo {author} {\bibfnamefont
  {M.}~\bibnamefont {Rispoli}}, \bibinfo {author} {\bibfnamefont
  {P.}~\bibnamefont {Zupancic}}, \bibinfo {author} {\bibfnamefont
  {Y.}~\bibnamefont {Lahini}}, \bibinfo {author} {\bibfnamefont
  {R.}~\bibnamefont {Islam}}, \ and\ \bibinfo {author} {\bibfnamefont
  {M.}~\bibnamefont {Greiner}},\ }\href {\doibase 10.1126/science.1260364}
  {\bibfield  {journal} {\bibinfo  {journal} {Science}\ }\textbf {\bibinfo
  {volume} {347}},\ \bibinfo {pages} {1229} (\bibinfo {year}
  {2015})}\BibitemShut {NoStop}%
\bibitem [{\citenamefont {Fukuhara}\ \emph {et~al.}(2015)\citenamefont
  {Fukuhara}, \citenamefont {Hild}, \citenamefont {Zeiher}, \citenamefont
  {Schau\ss{}}, \citenamefont {Bloch}, \citenamefont {Endres},\ and\
  \citenamefont {Gross}}]{FT15}%
  \BibitemOpen
  \bibfield  {author} {\bibinfo {author} {\bibfnamefont {T.}~\bibnamefont
  {Fukuhara}}, \bibinfo {author} {\bibfnamefont {S.}~\bibnamefont {Hild}},
  \bibinfo {author} {\bibfnamefont {J.}~\bibnamefont {Zeiher}}, \bibinfo
  {author} {\bibfnamefont {P.}~\bibnamefont {Schau\ss{}}}, \bibinfo {author}
  {\bibfnamefont {I.}~\bibnamefont {Bloch}}, \bibinfo {author} {\bibfnamefont
  {M.}~\bibnamefont {Endres}}, \ and\ \bibinfo {author} {\bibfnamefont
  {C.}~\bibnamefont {Gross}},\ }\href {\doibase 10.1103/PhysRevLett.115.035302}
  {\bibfield  {journal} {\bibinfo  {journal} {Phys. Rev. Lett.}\ }\textbf
  {\bibinfo {volume} {115}},\ \bibinfo {pages} {035302} (\bibinfo {year}
  {2015})}\BibitemShut {NoStop}%
\bibitem [{\citenamefont {Bloch}\ \emph {et~al.}(2012)\citenamefont {Bloch},
  \citenamefont {Dalibard},\ and\ \citenamefont {Nascimb\`ene}}]{BI12}%
  \BibitemOpen
  \bibfield  {author} {\bibinfo {author} {\bibfnamefont {I.}~\bibnamefont
  {Bloch}}, \bibinfo {author} {\bibfnamefont {J.}~\bibnamefont {Dalibard}}, \
  and\ \bibinfo {author} {\bibfnamefont {S.}~\bibnamefont {Nascimb\`ene}},\
  }\href@noop {} {\bibfield  {journal} {\bibinfo  {journal} {Nat. Phys.}\
  }\textbf {\bibinfo {volume} {8}},\ \bibinfo {pages} {267} (\bibinfo {year}
  {2012})}\BibitemShut {NoStop}%
\bibitem [{\citenamefont {Georgescu}\ \emph {et~al.}(2014)\citenamefont
  {Georgescu}, \citenamefont {Ashhab},\ and\ \citenamefont {Nori}}]{GIM14}%
  \BibitemOpen
  \bibfield  {author} {\bibinfo {author} {\bibfnamefont {I.~M.}\ \bibnamefont
  {Georgescu}}, \bibinfo {author} {\bibfnamefont {S.}~\bibnamefont {Ashhab}}, \
  and\ \bibinfo {author} {\bibfnamefont {F.}~\bibnamefont {Nori}},\ }\href
  {\doibase 10.1103/RevModPhys.86.153} {\bibfield  {journal} {\bibinfo
  {journal} {Rev. Mod. Phys.}\ }\textbf {\bibinfo {volume} {86}},\ \bibinfo
  {pages} {153} (\bibinfo {year} {2014})}\BibitemShut {NoStop}%
\bibitem [{\citenamefont {Nelson}\ \emph {et~al.}(2007)\citenamefont {Nelson},
  \citenamefont {Li},\ and\ \citenamefont {Weiss}}]{NKD07}%
  \BibitemOpen
  \bibfield  {author} {\bibinfo {author} {\bibfnamefont {K.~D.}\ \bibnamefont
  {Nelson}}, \bibinfo {author} {\bibfnamefont {X.}~\bibnamefont {Li}}, \ and\
  \bibinfo {author} {\bibfnamefont {D.~S.}\ \bibnamefont {Weiss}},\ }\href@noop
  {} {\bibfield  {journal} {\bibinfo  {journal} {Nat. Phys.}\ }\textbf
  {\bibinfo {volume} {3}},\ \bibinfo {pages} {556} (\bibinfo {year}
  {2007})}\BibitemShut {NoStop}%
\bibitem [{\citenamefont {Byrnes}\ \emph {et~al.}(2012)\citenamefont {Byrnes},
  \citenamefont {Wen},\ and\ \citenamefont {Yamamoto}}]{BT12}%
  \BibitemOpen
  \bibfield  {author} {\bibinfo {author} {\bibfnamefont {T.}~\bibnamefont
  {Byrnes}}, \bibinfo {author} {\bibfnamefont {K.}~\bibnamefont {Wen}}, \ and\
  \bibinfo {author} {\bibfnamefont {Y.}~\bibnamefont {Yamamoto}},\ }\href
  {\doibase 10.1103/PhysRevA.85.040306} {\bibfield  {journal} {\bibinfo
  {journal} {Phys. Rev. A}\ }\textbf {\bibinfo {volume} {85}},\ \bibinfo
  {pages} {040306} (\bibinfo {year} {2012})}\BibitemShut {NoStop}%
\bibitem [{\citenamefont {Daley}(2014)}]{DAJ14}%
  \BibitemOpen
  \bibfield  {author} {\bibinfo {author} {\bibfnamefont {A.~J.}\ \bibnamefont
  {Daley}},\ }\href@noop {} {\bibfield  {journal} {\bibinfo  {journal} {Adv.
  Phys.}\ }\textbf {\bibinfo {volume} {63}},\ \bibinfo {pages} {77} (\bibinfo
  {year} {2014})}\BibitemShut {NoStop}%
\bibitem [{\citenamefont {Patil}\ \emph {et~al.}(2014)\citenamefont {Patil},
  \citenamefont {Chakram}, \citenamefont {Aycock},\ and\ \citenamefont
  {Vengalattore}}]{PYS14}%
  \BibitemOpen
  \bibfield  {author} {\bibinfo {author} {\bibfnamefont {Y.~S.}\ \bibnamefont
  {Patil}}, \bibinfo {author} {\bibfnamefont {S.}~\bibnamefont {Chakram}},
  \bibinfo {author} {\bibfnamefont {L.~M.}\ \bibnamefont {Aycock}}, \ and\
  \bibinfo {author} {\bibfnamefont {M.}~\bibnamefont {Vengalattore}},\ }\href
  {\doibase 10.1103/PhysRevA.90.033422} {\bibfield  {journal} {\bibinfo
  {journal} {Phys. Rev. A}\ }\textbf {\bibinfo {volume} {90}},\ \bibinfo
  {pages} {033422} (\bibinfo {year} {2014})}\BibitemShut {NoStop}%
\bibitem [{\citenamefont {Ke\ss{}ler}\ \emph {et~al.}(2012)\citenamefont
  {Ke\ss{}ler}, \citenamefont {Holzner}, \citenamefont {McCulloch},
  \citenamefont {von Delft},\ and\ \citenamefont {Marquardt}}]{SK12}%
  \BibitemOpen
  \bibfield  {author} {\bibinfo {author} {\bibfnamefont {S.}~\bibnamefont
  {Ke\ss{}ler}}, \bibinfo {author} {\bibfnamefont {A.}~\bibnamefont {Holzner}},
  \bibinfo {author} {\bibfnamefont {I.~P.}\ \bibnamefont {McCulloch}}, \bibinfo
  {author} {\bibfnamefont {J.}~\bibnamefont {von Delft}}, \ and\ \bibinfo
  {author} {\bibfnamefont {F.}~\bibnamefont {Marquardt}},\ }\href {\doibase
  10.1103/PhysRevA.85.011605} {\bibfield  {journal} {\bibinfo  {journal} {Phys.
  Rev. A}\ }\textbf {\bibinfo {volume} {85}},\ \bibinfo {pages} {011605}
  (\bibinfo {year} {2012})}\BibitemShut {NoStop}%
\bibitem [{\citenamefont {Daley}\ \emph {et~al.}(2012)\citenamefont {Daley},
  \citenamefont {Pichler}, \citenamefont {Schachenmayer},\ and\ \citenamefont
  {Zoller}}]{DAJ12}%
  \BibitemOpen
  \bibfield  {author} {\bibinfo {author} {\bibfnamefont {A.~J.}\ \bibnamefont
  {Daley}}, \bibinfo {author} {\bibfnamefont {H.}~\bibnamefont {Pichler}},
  \bibinfo {author} {\bibfnamefont {J.}~\bibnamefont {Schachenmayer}}, \ and\
  \bibinfo {author} {\bibfnamefont {P.}~\bibnamefont {Zoller}},\ }\href
  {\doibase 10.1103/PhysRevLett.109.020505} {\bibfield  {journal} {\bibinfo
  {journal} {Phys. Rev. Lett.}\ }\textbf {\bibinfo {volume} {109}},\ \bibinfo
  {pages} {020505} (\bibinfo {year} {2012})}\BibitemShut {NoStop}%
\bibitem [{\citenamefont {Douglas}\ and\ \citenamefont
  {Burnett}(2012)}]{DJS12}%
  \BibitemOpen
  \bibfield  {author} {\bibinfo {author} {\bibfnamefont {J.~S.}\ \bibnamefont
  {Douglas}}\ and\ \bibinfo {author} {\bibfnamefont {K.}~\bibnamefont
  {Burnett}},\ }\href {\doibase 10.1103/PhysRevA.86.052120} {\bibfield
  {journal} {\bibinfo  {journal} {Phys. Rev. A}\ }\textbf {\bibinfo {volume}
  {86}},\ \bibinfo {pages} {052120} (\bibinfo {year} {2012})}\BibitemShut
  {NoStop}%
\bibitem [{\citenamefont {Ke\ssŸler}\ \emph {et~al.}(2013)\citenamefont
  {Ke\ssŸler}, \citenamefont {McCulloch},\ and\ \citenamefont
  {Marquardt}}]{SK13}%
  \BibitemOpen
  \bibfield  {author} {\bibinfo {author} {\bibfnamefont {S.}~\bibnamefont
  {Ke\ssŸler}}, \bibinfo {author} {\bibfnamefont {I.~P.}\ \bibnamefont
  {McCulloch}}, \ and\ \bibinfo {author} {\bibfnamefont {F.}~\bibnamefont
  {Marquardt}},\ }\href {http://stacks.iop.org/1367-2630/15/i=5/a=053043}
  {\bibfield  {journal} {\bibinfo  {journal} {New J. Phys.}\ }\textbf {\bibinfo
  {volume} {15}},\ \bibinfo {pages} {053043} (\bibinfo {year}
  {2013})}\BibitemShut {NoStop}%
\bibitem [{\citenamefont {{Patil}}\ \emph {et~al.}(2014)\citenamefont
  {{Patil}}, \citenamefont {{Chakram}},\ and\ \citenamefont
  {{Vengalattore}}}]{PYS142}%
  \BibitemOpen
  \bibfield  {author} {\bibinfo {author} {\bibfnamefont {Y.~S.}\ \bibnamefont
  {{Patil}}}, \bibinfo {author} {\bibfnamefont {S.}~\bibnamefont {{Chakram}}},
  \ and\ \bibinfo {author} {\bibfnamefont {M.}~\bibnamefont {{Vengalattore}}},\
  }\href@noop {} {\bibfield  {journal} {\bibinfo  {journal} {ArXiv e-prints}\ }
  (\bibinfo {year} {2014})},\ \Eprint {http://arxiv.org/abs/1411.2678}
  {arXiv:1411.2678 [cond-mat.quant-gas]} \BibitemShut {NoStop}%
\bibitem [{\citenamefont {Wiseman}\ and\ \citenamefont
  {Milburn}(2010)}]{WHM10}%
  \BibitemOpen
  \bibfield  {author} {\bibinfo {author} {\bibfnamefont {H.}~\bibnamefont
  {Wiseman}}\ and\ \bibinfo {author} {\bibfnamefont {G.}~\bibnamefont
  {Milburn}},\ }\href {http://books.google.co.uk/books?id=ZNjvHaH8qA4C} {\emph
  {\bibinfo {title} {Quantum Measurement and Control}}}\ (\bibinfo  {publisher}
  {Cambridge University Press},\ \bibinfo {year} {2010})\BibitemShut {NoStop}%
\bibitem [{\citenamefont {{Sayrin}}\ \emph {et~al.}(2011)\citenamefont
  {{Sayrin}}, \citenamefont {{Dotsenko}}, \citenamefont {{Zhou}}, \citenamefont
  {{Peaudecerf}}, \citenamefont {{Rybarczyk}}, \citenamefont {{Gleyzes}},
  \citenamefont {{Rouchon}}, \citenamefont {{Mirrahimi}}, \citenamefont
  {{Amini}}, \citenamefont {{Brune}}, \citenamefont {{Raimond}},\ and\
  \citenamefont {{Haroche}}}]{SC11}%
  \BibitemOpen
  \bibfield  {author} {\bibinfo {author} {\bibfnamefont {C.}~\bibnamefont
  {{Sayrin}}}, \bibinfo {author} {\bibfnamefont {I.}~\bibnamefont
  {{Dotsenko}}}, \bibinfo {author} {\bibfnamefont {X.}~\bibnamefont {{Zhou}}},
  \bibinfo {author} {\bibfnamefont {B.}~\bibnamefont {{Peaudecerf}}}, \bibinfo
  {author} {\bibfnamefont {T.}~\bibnamefont {{Rybarczyk}}}, \bibinfo {author}
  {\bibfnamefont {S.}~\bibnamefont {{Gleyzes}}}, \bibinfo {author}
  {\bibfnamefont {P.}~\bibnamefont {{Rouchon}}}, \bibinfo {author}
  {\bibfnamefont {M.}~\bibnamefont {{Mirrahimi}}}, \bibinfo {author}
  {\bibfnamefont {H.}~\bibnamefont {{Amini}}}, \bibinfo {author} {\bibfnamefont
  {M.}~\bibnamefont {{Brune}}}, \bibinfo {author} {\bibfnamefont {J.-M.}\
  \bibnamefont {{Raimond}}}, \ and\ \bibinfo {author} {\bibfnamefont
  {S.}~\bibnamefont {{Haroche}}},\ }\href@noop {} {\bibfield  {journal}
  {\bibinfo  {journal} {Nature}\ }\textbf {\bibinfo {volume} {477}},\ \bibinfo
  {pages} {73} (\bibinfo {year} {2011})}\BibitemShut {NoStop}%
\bibitem [{\citenamefont {Brune}\ \emph {et~al.}(1990)\citenamefont {Brune},
  \citenamefont {Haroche}, \citenamefont {Lefevre}, \citenamefont {Raimond},\
  and\ \citenamefont {Zagury}}]{BM90}%
  \BibitemOpen
  \bibfield  {author} {\bibinfo {author} {\bibfnamefont {M.}~\bibnamefont
  {Brune}}, \bibinfo {author} {\bibfnamefont {S.}~\bibnamefont {Haroche}},
  \bibinfo {author} {\bibfnamefont {V.}~\bibnamefont {Lefevre}}, \bibinfo
  {author} {\bibfnamefont {J.~M.}\ \bibnamefont {Raimond}}, \ and\ \bibinfo
  {author} {\bibfnamefont {N.}~\bibnamefont {Zagury}},\ }\href {\doibase
  10.1103/PhysRevLett.65.976} {\bibfield  {journal} {\bibinfo  {journal} {Phys.
  Rev. Lett.}\ }\textbf {\bibinfo {volume} {65}},\ \bibinfo {pages} {976}
  (\bibinfo {year} {1990})}\BibitemShut {NoStop}%
\bibitem [{\citenamefont {Guerlin}\ \emph {et~al.}(2007)\citenamefont
  {Guerlin}, \citenamefont {Bernu}, \citenamefont {Del\`eglise}, \citenamefont
  {Sayrin}, \citenamefont {Gleyzes}, \citenamefont {Kuhr}, \citenamefont
  {Brune}, \citenamefont {Raimond},\ and\ \citenamefont {Haroche}}]{GC07}%
  \BibitemOpen
  \bibfield  {author} {\bibinfo {author} {\bibfnamefont {C.}~\bibnamefont
  {Guerlin}}, \bibinfo {author} {\bibfnamefont {J.}~\bibnamefont {Bernu}},
  \bibinfo {author} {\bibfnamefont {S.}~\bibnamefont {Del\`eglise}}, \bibinfo
  {author} {\bibfnamefont {C.}~\bibnamefont {Sayrin}}, \bibinfo {author}
  {\bibfnamefont {S.}~\bibnamefont {Gleyzes}}, \bibinfo {author} {\bibfnamefont
  {S.}~\bibnamefont {Kuhr}}, \bibinfo {author} {\bibfnamefont {M.}~\bibnamefont
  {Brune}}, \bibinfo {author} {\bibfnamefont {J.-M.}\ \bibnamefont {Raimond}},
  \ and\ \bibinfo {author} {\bibfnamefont {S.}~\bibnamefont {Haroche}},\
  }\href@noop {} {\bibfield  {journal} {\bibinfo  {journal} {Nature}\ }\textbf
  {\bibinfo {volume} {448}},\ \bibinfo {pages} {889} (\bibinfo {year}
  {2007})}\BibitemShut {NoStop}%
\bibitem [{\citenamefont {{Rybarczyk}}\ \emph {et~al.}(2014)\citenamefont
  {{Rybarczyk}}, \citenamefont {{Gerlich}}, \citenamefont {{Peaudecerf}},
  \citenamefont {{Penasa}}, \citenamefont {{Julsgaard}}, \citenamefont
  {{Moelmer}}, \citenamefont {{Gleyzes}}, \citenamefont {{Brune}},
  \citenamefont {{Raimond}}, \citenamefont {{Haroche}},\ and\ \citenamefont
  {{Dotsenko}}}]{RT14}%
  \BibitemOpen
  \bibfield  {author} {\bibinfo {author} {\bibfnamefont {T.}~\bibnamefont
  {{Rybarczyk}}}, \bibinfo {author} {\bibfnamefont {S.}~\bibnamefont
  {{Gerlich}}}, \bibinfo {author} {\bibfnamefont {B.}~\bibnamefont
  {{Peaudecerf}}}, \bibinfo {author} {\bibfnamefont {M.}~\bibnamefont
  {{Penasa}}}, \bibinfo {author} {\bibfnamefont {B.}~\bibnamefont
  {{Julsgaard}}}, \bibinfo {author} {\bibfnamefont {K.}~\bibnamefont
  {{Moelmer}}}, \bibinfo {author} {\bibfnamefont {S.}~\bibnamefont
  {{Gleyzes}}}, \bibinfo {author} {\bibfnamefont {M.}~\bibnamefont {{Brune}}},
  \bibinfo {author} {\bibfnamefont {J.-M.}\ \bibnamefont {{Raimond}}}, \bibinfo
  {author} {\bibfnamefont {S.}~\bibnamefont {{Haroche}}}, \ and\ \bibinfo
  {author} {\bibfnamefont {I.}~\bibnamefont {{Dotsenko}}},\ }\href@noop {}
  {\bibfield  {journal} {\bibinfo  {journal} {ArXiv e-prints}\ } (\bibinfo
  {year} {2014})},\ \Eprint {http://arxiv.org/abs/1409.0958} {arXiv:1409.0958
  [quant-ph]} \BibitemShut {NoStop}%
\bibitem [{\citenamefont {Holland}\ \emph {et~al.}(1996)\citenamefont
  {Holland}, \citenamefont {Marksteiner}, \citenamefont {Marte},\ and\
  \citenamefont {Zoller}}]{HM96}%
  \BibitemOpen
  \bibfield  {author} {\bibinfo {author} {\bibfnamefont {M.}~\bibnamefont
  {Holland}}, \bibinfo {author} {\bibfnamefont {S.}~\bibnamefont
  {Marksteiner}}, \bibinfo {author} {\bibfnamefont {P.}~\bibnamefont {Marte}},
  \ and\ \bibinfo {author} {\bibfnamefont {P.}~\bibnamefont {Zoller}},\ }\href
  {\doibase 10.1103/PhysRevLett.76.3683} {\bibfield  {journal} {\bibinfo
  {journal} {Phys. Rev. Lett.}\ }\textbf {\bibinfo {volume} {76}},\ \bibinfo
  {pages} {3683} (\bibinfo {year} {1996})}\BibitemShut {NoStop}%
\bibitem [{\citenamefont {Javanainen}(1995)}]{JJ95}%
  \BibitemOpen
  \bibfield  {author} {\bibinfo {author} {\bibfnamefont {J.}~\bibnamefont
  {Javanainen}},\ }\href {\doibase 10.1103/PhysRevLett.75.1927} {\bibfield
  {journal} {\bibinfo  {journal} {Phys. Rev. Lett.}\ }\textbf {\bibinfo
  {volume} {75}},\ \bibinfo {pages} {1927} (\bibinfo {year}
  {1995})}\BibitemShut {NoStop}%
\bibitem [{\citenamefont {Javanainen}\ and\ \citenamefont
  {Ruostekoski}(1995)}]{JJ952}%
  \BibitemOpen
  \bibfield  {author} {\bibinfo {author} {\bibfnamefont {J.}~\bibnamefont
  {Javanainen}}\ and\ \bibinfo {author} {\bibfnamefont {J.}~\bibnamefont
  {Ruostekoski}},\ }\href {\doibase 10.1103/PhysRevA.52.3033} {\bibfield
  {journal} {\bibinfo  {journal} {Phys. Rev. A}\ }\textbf {\bibinfo {volume}
  {52}},\ \bibinfo {pages} {3033} (\bibinfo {year} {1995})}\BibitemShut
  {NoStop}%
\bibitem [{\citenamefont {Idziaszek}\ \emph {et~al.}(2000)\citenamefont
  {Idziaszek}, \citenamefont {Rza\ifmmode \mbox{\c{}}\else
  \c{}\fi{}\ifmmode~\dot{z}\else \.{z}\fi{}ewski},\ and\ \citenamefont
  {Lewenstein}}]{IZ00}%
  \BibitemOpen
  \bibfield  {author} {\bibinfo {author} {\bibfnamefont {Z.}~\bibnamefont
  {Idziaszek}}, \bibinfo {author} {\bibfnamefont {K.}~\bibnamefont {Rza\ifmmode
  \mbox{\c{}}\else \c{}\fi{}\ifmmode~\dot{z}\else \.{z}\fi{}ewski}}, \ and\
  \bibinfo {author} {\bibfnamefont {M.}~\bibnamefont {Lewenstein}},\ }\href
  {\doibase 10.1103/PhysRevA.61.053608} {\bibfield  {journal} {\bibinfo
  {journal} {Phys. Rev. A}\ }\textbf {\bibinfo {volume} {61}},\ \bibinfo
  {pages} {053608} (\bibinfo {year} {2000})}\BibitemShut {NoStop}%
\bibitem [{\citenamefont {Rey}\ \emph {et~al.}(2005)\citenamefont {Rey},
  \citenamefont {Blakie}, \citenamefont {Pupillo}, \citenamefont {Williams},\
  and\ \citenamefont {Clark}}]{RAM05}%
  \BibitemOpen
  \bibfield  {author} {\bibinfo {author} {\bibfnamefont {A.~M.}\ \bibnamefont
  {Rey}}, \bibinfo {author} {\bibfnamefont {P.~B.}\ \bibnamefont {Blakie}},
  \bibinfo {author} {\bibfnamefont {G.}~\bibnamefont {Pupillo}}, \bibinfo
  {author} {\bibfnamefont {C.~J.}\ \bibnamefont {Williams}}, \ and\ \bibinfo
  {author} {\bibfnamefont {C.~W.}\ \bibnamefont {Clark}},\ }\href {\doibase
  10.1103/PhysRevA.72.023407} {\bibfield  {journal} {\bibinfo  {journal} {Phys.
  Rev. A}\ }\textbf {\bibinfo {volume} {72}},\ \bibinfo {pages} {023407}
  (\bibinfo {year} {2005})}\BibitemShut {NoStop}%
\bibitem [{\citenamefont {\L{}akomy}\ \emph {et~al.}(2009)\citenamefont
  {\L{}akomy}, \citenamefont {Idziaszek},\ and\ \citenamefont
  {Trippenbach}}]{LK09}%
  \BibitemOpen
  \bibfield  {author} {\bibinfo {author} {\bibfnamefont {K.}~\bibnamefont
  {\L{}akomy}}, \bibinfo {author} {\bibfnamefont {Z.}~\bibnamefont
  {Idziaszek}}, \ and\ \bibinfo {author} {\bibfnamefont {M.}~\bibnamefont
  {Trippenbach}},\ }\href {\doibase 10.1103/PhysRevA.80.043404} {\bibfield
  {journal} {\bibinfo  {journal} {Phys. Rev. A}\ }\textbf {\bibinfo {volume}
  {80}},\ \bibinfo {pages} {043404} (\bibinfo {year} {2009})}\BibitemShut
  {NoStop}%
\bibitem [{\citenamefont {Douglas}\ and\ \citenamefont
  {Burnett}(2011)}]{DJS11}%
  \BibitemOpen
  \bibfield  {author} {\bibinfo {author} {\bibfnamefont {J.~S.}\ \bibnamefont
  {Douglas}}\ and\ \bibinfo {author} {\bibfnamefont {K.}~\bibnamefont
  {Burnett}},\ }\href {\doibase 10.1103/PhysRevA.84.033637} {\bibfield
  {journal} {\bibinfo  {journal} {Phys. Rev. A}\ }\textbf {\bibinfo {volume}
  {84}},\ \bibinfo {pages} {033637} (\bibinfo {year} {2011})}\BibitemShut
  {NoStop}%
\bibitem [{\citenamefont {Mekhov}\ \emph {et~al.}(2007)\citenamefont {Mekhov},
  \citenamefont {Maschler},\ and\ \citenamefont {Ritsch}}]{MIB07}%
  \BibitemOpen
  \bibfield  {author} {\bibinfo {author} {\bibfnamefont {I.~B.}\ \bibnamefont
  {Mekhov}}, \bibinfo {author} {\bibfnamefont {C.}~\bibnamefont {Maschler}}, \
  and\ \bibinfo {author} {\bibfnamefont {H.}~\bibnamefont {Ritsch}},\
  }\href@noop {} {\bibfield  {journal} {\bibinfo  {journal} {Nat. Phys.}\
  }\textbf {\bibinfo {volume} {3}},\ \bibinfo {pages} {319} (\bibinfo {year}
  {2007})}\BibitemShut {NoStop}%
\bibitem [{\citenamefont {Mekhov}\ and\ \citenamefont {Ritsch}(2009)}]{MIB09}%
  \BibitemOpen
  \bibfield  {author} {\bibinfo {author} {\bibfnamefont {I.~B.}\ \bibnamefont
  {Mekhov}}\ and\ \bibinfo {author} {\bibfnamefont {H.}~\bibnamefont
  {Ritsch}},\ }\href {\doibase 10.1103/PhysRevLett.102.020403} {\bibfield
  {journal} {\bibinfo  {journal} {Phys. Rev. Lett.}\ }\textbf {\bibinfo
  {volume} {102}},\ \bibinfo {pages} {020403} (\bibinfo {year}
  {2009})}\BibitemShut {NoStop}%
\bibitem [{\citenamefont {Cheuk}\ \emph {et~al.}(2015)\citenamefont {Cheuk},
  \citenamefont {Nichols}, \citenamefont {Okan}, \citenamefont {Gersdorf},
  \citenamefont {Ramasesh}, \citenamefont {Bakr}, \citenamefont {Lompe},\ and\
  \citenamefont {Zwierlein}}]{CLW15}%
  \BibitemOpen
  \bibfield  {author} {\bibinfo {author} {\bibfnamefont {L.~W.}\ \bibnamefont
  {Cheuk}}, \bibinfo {author} {\bibfnamefont {M.~A.}\ \bibnamefont {Nichols}},
  \bibinfo {author} {\bibfnamefont {M.}~\bibnamefont {Okan}}, \bibinfo {author}
  {\bibfnamefont {T.}~\bibnamefont {Gersdorf}}, \bibinfo {author}
  {\bibfnamefont {V.~V.}\ \bibnamefont {Ramasesh}}, \bibinfo {author}
  {\bibfnamefont {W.~S.}\ \bibnamefont {Bakr}}, \bibinfo {author}
  {\bibfnamefont {T.}~\bibnamefont {Lompe}}, \ and\ \bibinfo {author}
  {\bibfnamefont {M.~W.}\ \bibnamefont {Zwierlein}},\ }\href {\doibase
  10.1103/PhysRevLett.114.193001} {\bibfield  {journal} {\bibinfo  {journal}
  {Phys. Rev. Lett.}\ }\textbf {\bibinfo {volume} {114}},\ \bibinfo {pages}
  {193001} (\bibinfo {year} {2015})}\BibitemShut {NoStop}%
\bibitem [{\citenamefont {Parsons}\ \emph {et~al.}(2015)\citenamefont
  {Parsons}, \citenamefont {Huber}, \citenamefont {Mazurenko}, \citenamefont
  {Chiu}, \citenamefont {Setiawan}, \citenamefont {Wooley-Brown}, \citenamefont
  {Blatt},\ and\ \citenamefont {Greiner}}]{PMF15}%
  \BibitemOpen
  \bibfield  {author} {\bibinfo {author} {\bibfnamefont {M.~F.}\ \bibnamefont
  {Parsons}}, \bibinfo {author} {\bibfnamefont {F.}~\bibnamefont {Huber}},
  \bibinfo {author} {\bibfnamefont {A.}~\bibnamefont {Mazurenko}}, \bibinfo
  {author} {\bibfnamefont {C.~S.}\ \bibnamefont {Chiu}}, \bibinfo {author}
  {\bibfnamefont {W.}~\bibnamefont {Setiawan}}, \bibinfo {author}
  {\bibfnamefont {K.}~\bibnamefont {Wooley-Brown}}, \bibinfo {author}
  {\bibfnamefont {S.}~\bibnamefont {Blatt}}, \ and\ \bibinfo {author}
  {\bibfnamefont {M.}~\bibnamefont {Greiner}},\ }\href {\doibase
  10.1103/PhysRevLett.114.213002} {\bibfield  {journal} {\bibinfo  {journal}
  {Phys. Rev. Lett.}\ }\textbf {\bibinfo {volume} {114}},\ \bibinfo {pages}
  {213002} (\bibinfo {year} {2015})}\BibitemShut {NoStop}%
\bibitem [{\citenamefont {Haller}\ \emph {et~al.}()\citenamefont {Haller},
  \citenamefont {Hudson}, \citenamefont {Kelly}, \citenamefont {Cotta},
  \citenamefont {Bruno}, \citenamefont {Bruce},\ and\ \citenamefont
  {Kuhr}}]{EH15}%
  \BibitemOpen
  \bibfield  {author} {\bibinfo {author} {\bibfnamefont {E.}~\bibnamefont
  {Haller}}, \bibinfo {author} {\bibfnamefont {J.}~\bibnamefont {Hudson}},
  \bibinfo {author} {\bibfnamefont {A.}~\bibnamefont {Kelly}}, \bibinfo
  {author} {\bibfnamefont {D.~A.}\ \bibnamefont {Cotta}}, \bibinfo {author}
  {\bibfnamefont {P.}~\bibnamefont {Bruno}}, \bibinfo {author} {\bibfnamefont
  {G.~D.}\ \bibnamefont {Bruce}}, \ and\ \bibinfo {author} {\bibfnamefont
  {S.}~\bibnamefont {Kuhr}},\ }\href {\doibase 10.1038/nphys3403} {\bibfield
  {journal} {\bibinfo  {journal} {Nat. Phys.}\ }10.1038/nphys3403}\BibitemShut
  {NoStop}%
\bibitem [{\citenamefont {Leibfried}\ \emph {et~al.}(2003)\citenamefont
  {Leibfried}, \citenamefont {Blatt}, \citenamefont {Monroe},\ and\
  \citenamefont {Wineland}}]{LD03}%
  \BibitemOpen
  \bibfield  {author} {\bibinfo {author} {\bibfnamefont {D.}~\bibnamefont
  {Leibfried}}, \bibinfo {author} {\bibfnamefont {R.}~\bibnamefont {Blatt}},
  \bibinfo {author} {\bibfnamefont {C.}~\bibnamefont {Monroe}}, \ and\ \bibinfo
  {author} {\bibfnamefont {D.}~\bibnamefont {Wineland}},\ }\href {\doibase
  10.1103/RevModPhys.75.281} {\bibfield  {journal} {\bibinfo  {journal} {Rev.
  Mod. Phys.}\ }\textbf {\bibinfo {volume} {75}},\ \bibinfo {pages} {281}
  (\bibinfo {year} {2003})}\BibitemShut {NoStop}%
\bibitem [{\citenamefont {Streed}\ \emph {et~al.}(2011)\citenamefont {Streed},
  \citenamefont {Norton}, \citenamefont {Jechow}, \citenamefont {Weinhold},\
  and\ \citenamefont {Kielpinski}}]{SEW11}%
  \BibitemOpen
  \bibfield  {author} {\bibinfo {author} {\bibfnamefont {E.~W.}\ \bibnamefont
  {Streed}}, \bibinfo {author} {\bibfnamefont {B.~G.}\ \bibnamefont {Norton}},
  \bibinfo {author} {\bibfnamefont {A.}~\bibnamefont {Jechow}}, \bibinfo
  {author} {\bibfnamefont {T.~J.}\ \bibnamefont {Weinhold}}, \ and\ \bibinfo
  {author} {\bibfnamefont {D.}~\bibnamefont {Kielpinski}},\ }\href {\doibase
  10.1103/PhysRevLett.106.010502} {\bibfield  {journal} {\bibinfo  {journal}
  {Phys. Rev. Lett.}\ }\textbf {\bibinfo {volume} {106}},\ \bibinfo {pages}
  {010502} (\bibinfo {year} {2011})}\BibitemShut {NoStop}%
\bibitem [{\citenamefont {Jechow}\ \emph {et~al.}(2011)\citenamefont {Jechow},
  \citenamefont {Streed}, \citenamefont {Norton}, \citenamefont {Petrasiunas},\
  and\ \citenamefont {Kielpinski}}]{AJ11}%
  \BibitemOpen
  \bibfield  {author} {\bibinfo {author} {\bibfnamefont {A.}~\bibnamefont
  {Jechow}}, \bibinfo {author} {\bibfnamefont {E.~W.}\ \bibnamefont {Streed}},
  \bibinfo {author} {\bibfnamefont {B.~G.}\ \bibnamefont {Norton}}, \bibinfo
  {author} {\bibfnamefont {M.~J.}\ \bibnamefont {Petrasiunas}}, \ and\ \bibinfo
  {author} {\bibfnamefont {D.}~\bibnamefont {Kielpinski}},\ }\href {\doibase
  10.1364/OL.36.001371} {\bibfield  {journal} {\bibinfo  {journal} {Opt.
  Lett.}\ }\textbf {\bibinfo {volume} {36}},\ \bibinfo {pages} {1371} (\bibinfo
  {year} {2011})}\BibitemShut {NoStop}%
\bibitem [{\citenamefont {Moerner}\ and\ \citenamefont {Kador}(1989)}]{MWE89}%
  \BibitemOpen
  \bibfield  {author} {\bibinfo {author} {\bibfnamefont {W.~E.}\ \bibnamefont
  {Moerner}}\ and\ \bibinfo {author} {\bibfnamefont {L.}~\bibnamefont
  {Kador}},\ }\href {\doibase 10.1103/PhysRevLett.62.2535} {\bibfield
  {journal} {\bibinfo  {journal} {Phys. Rev. Lett.}\ }\textbf {\bibinfo
  {volume} {62}},\ \bibinfo {pages} {2535} (\bibinfo {year}
  {1989})}\BibitemShut {NoStop}%
\bibitem [{\citenamefont {Dickson}\ \emph {et~al.}(1997)\citenamefont
  {Dickson}, \citenamefont {Cubitt}, \citenamefont {Tsien},\ and\ \citenamefont
  {Moerner}}]{DRM97}%
  \BibitemOpen
  \bibfield  {author} {\bibinfo {author} {\bibfnamefont {R.~M.}\ \bibnamefont
  {Dickson}}, \bibinfo {author} {\bibfnamefont {A.~B.}\ \bibnamefont {Cubitt}},
  \bibinfo {author} {\bibfnamefont {R.~Y.}\ \bibnamefont {Tsien}}, \ and\
  \bibinfo {author} {\bibfnamefont {W.~E.}\ \bibnamefont {Moerner}},\
  }\href@noop {} {\bibfield  {journal} {\bibinfo  {journal} {Nature}\ }\textbf
  {\bibinfo {volume} {388}},\ \bibinfo {pages} {355} (\bibinfo {year}
  {1997})}\BibitemShut {NoStop}%
\bibitem [{\citenamefont {Betzig}\ \emph {et~al.}(2006)\citenamefont {Betzig},
  \citenamefont {Patterson}, \citenamefont {Sougrat}, \citenamefont
  {Lindwasser}, \citenamefont {Olenych}, \citenamefont {Bonifacino},
  \citenamefont {Davidson}, \citenamefont {Lippincott-Schwartz},\ and\
  \citenamefont {Hess}}]{BE06}%
  \BibitemOpen
  \bibfield  {author} {\bibinfo {author} {\bibfnamefont {E.}~\bibnamefont
  {Betzig}}, \bibinfo {author} {\bibfnamefont {G.~H.}\ \bibnamefont
  {Patterson}}, \bibinfo {author} {\bibfnamefont {R.}~\bibnamefont {Sougrat}},
  \bibinfo {author} {\bibfnamefont {O.~W.}\ \bibnamefont {Lindwasser}},
  \bibinfo {author} {\bibfnamefont {S.}~\bibnamefont {Olenych}}, \bibinfo
  {author} {\bibfnamefont {J.~S.}\ \bibnamefont {Bonifacino}}, \bibinfo
  {author} {\bibfnamefont {M.~W.}\ \bibnamefont {Davidson}}, \bibinfo {author}
  {\bibfnamefont {J.}~\bibnamefont {Lippincott-Schwartz}}, \ and\ \bibinfo
  {author} {\bibfnamefont {H.~F.}\ \bibnamefont {Hess}},\ }\href@noop {}
  {\bibfield  {journal} {\bibinfo  {journal} {Science}\ }\textbf {\bibinfo
  {volume} {313}},\ \bibinfo {pages} {1642} (\bibinfo {year}
  {2006})}\BibitemShut {NoStop}%
\bibitem [{\citenamefont {Rust}\ \emph {et~al.}(2006)\citenamefont {Rust},
  \citenamefont {Bates},\ and\ \citenamefont {Zhuang}}]{RMJ06}%
  \BibitemOpen
  \bibfield  {author} {\bibinfo {author} {\bibfnamefont {M.~J.}\ \bibnamefont
  {Rust}}, \bibinfo {author} {\bibfnamefont {M.}~\bibnamefont {Bates}}, \ and\
  \bibinfo {author} {\bibfnamefont {X.}~\bibnamefont {Zhuang}},\ }\href@noop {}
  {\bibfield  {journal} {\bibinfo  {journal} {Nat. Meth.}\ }\textbf {\bibinfo
  {volume} {3}},\ \bibinfo {pages} {793} (\bibinfo {year} {2006})}\BibitemShut
  {NoStop}%
\bibitem [{\citenamefont {Sahl}\ and\ \citenamefont {Moerner}(2013)}]{SJS13}%
  \BibitemOpen
  \bibfield  {author} {\bibinfo {author} {\bibfnamefont {S.~J.}\ \bibnamefont
  {Sahl}}\ and\ \bibinfo {author} {\bibfnamefont {W.}~\bibnamefont {Moerner}},\
  }\href@noop {} {\bibfield  {journal} {\bibinfo  {journal} {Curr. Opin.
  Struct. Biol.}\ }\textbf {\bibinfo {volume} {23}},\ \bibinfo {pages} {778 }
  (\bibinfo {year} {2013})}\BibitemShut {NoStop}%
\bibitem [{\citenamefont {{Ashida}}\ and\ \citenamefont {{Ueda}}(2015)}]{YA15}%
  \BibitemOpen
  \bibfield  {author} {\bibinfo {author} {\bibfnamefont {Y.}~\bibnamefont
  {{Ashida}}}\ and\ \bibinfo {author} {\bibfnamefont {M.}~\bibnamefont
  {{Ueda}}},\ }\href@noop {} {\bibfield  {journal} {\bibinfo  {journal} {ArXiv
  e-prints}\ } (\bibinfo {year} {2015})},\ \Eprint
  {http://arxiv.org/abs/1505.00507} {arXiv:1505.00507 [physics.optics]}
  \BibitemShut {NoStop}%
\bibitem [{\citenamefont {Gullans}\ \emph {et~al.}(2012)\citenamefont
  {Gullans}, \citenamefont {Tiecke}, \citenamefont {Chang}, \citenamefont
  {Feist}, \citenamefont {Thompson}, \citenamefont {Cirac}, \citenamefont
  {Zoller},\ and\ \citenamefont {Lukin}}]{GM12}%
  \BibitemOpen
  \bibfield  {author} {\bibinfo {author} {\bibfnamefont {M.}~\bibnamefont
  {Gullans}}, \bibinfo {author} {\bibfnamefont {T.~G.}\ \bibnamefont {Tiecke}},
  \bibinfo {author} {\bibfnamefont {D.~E.}\ \bibnamefont {Chang}}, \bibinfo
  {author} {\bibfnamefont {J.}~\bibnamefont {Feist}}, \bibinfo {author}
  {\bibfnamefont {J.~D.}\ \bibnamefont {Thompson}}, \bibinfo {author}
  {\bibfnamefont {J.~I.}\ \bibnamefont {Cirac}}, \bibinfo {author}
  {\bibfnamefont {P.}~\bibnamefont {Zoller}}, \ and\ \bibinfo {author}
  {\bibfnamefont {M.~D.}\ \bibnamefont {Lukin}},\ }\href {\doibase
  10.1103/PhysRevLett.109.235309} {\bibfield  {journal} {\bibinfo  {journal}
  {Phys. Rev. Lett.}\ }\textbf {\bibinfo {volume} {109}},\ \bibinfo {pages}
  {235309} (\bibinfo {year} {2012})}\BibitemShut {NoStop}%
\bibitem [{\citenamefont {Gonz\'alez-Tudela}\ \emph {et~al.}(2015)\citenamefont
  {Gonz\'alez-Tudela}, \citenamefont {Hung}, \citenamefont {Chang},
  \citenamefont {Cirac},\ and\ \citenamefont {Kimble}}]{AGT15}%
  \BibitemOpen
  \bibfield  {author} {\bibinfo {author} {\bibfnamefont {A.}~\bibnamefont
  {Gonz\'alez-Tudela}}, \bibinfo {author} {\bibfnamefont {C.-L.}\ \bibnamefont
  {Hung}}, \bibinfo {author} {\bibfnamefont {D.~E.}\ \bibnamefont {Chang}},
  \bibinfo {author} {\bibfnamefont {J.~I.}\ \bibnamefont {Cirac}}, \ and\
  \bibinfo {author} {\bibfnamefont {H.~J.}\ \bibnamefont {Kimble}},\
  }\href@noop {} {\bibfield  {journal} {\bibinfo  {journal} {Nat. Photo.}\
  }\textbf {\bibinfo {volume} {9}},\ \bibinfo {pages} {320} (\bibinfo {year}
  {2015})}\BibitemShut {NoStop}%
\bibitem [{\citenamefont {{Nascimbene}}\ \emph {et~al.}(2015)\citenamefont
  {{Nascimbene}}, \citenamefont {{Goldman}}, \citenamefont {{Cooper}},\ and\
  \citenamefont {{Dalibard}}}]{SN15}%
  \BibitemOpen
  \bibfield  {author} {\bibinfo {author} {\bibfnamefont {S.}~\bibnamefont
  {{Nascimbene}}}, \bibinfo {author} {\bibfnamefont {N.}~\bibnamefont
  {{Goldman}}}, \bibinfo {author} {\bibfnamefont {N.~R.}\ \bibnamefont
  {{Cooper}}}, \ and\ \bibinfo {author} {\bibfnamefont {J.}~\bibnamefont
  {{Dalibard}}},\ }\href@noop {} {\bibfield  {journal} {\bibinfo  {journal}
  {ArXiv e-prints}\ } (\bibinfo {year} {2015})},\ \Eprint
  {http://arxiv.org/abs/1506.00558} {arXiv:1506.00558 [cond-mat.quant-gas]}
  \BibitemShut {NoStop}%
\bibitem [{\citenamefont {Klar}\ \emph {et~al.}(2000)\citenamefont {Klar},
  \citenamefont {Jakobs}, \citenamefont {Dyba}, \citenamefont {Egner},\ and\
  \citenamefont {Hell}}]{KTA00}%
  \BibitemOpen
  \bibfield  {author} {\bibinfo {author} {\bibfnamefont {T.~A.}\ \bibnamefont
  {Klar}}, \bibinfo {author} {\bibfnamefont {S.}~\bibnamefont {Jakobs}},
  \bibinfo {author} {\bibfnamefont {M.}~\bibnamefont {Dyba}}, \bibinfo {author}
  {\bibfnamefont {A.}~\bibnamefont {Egner}}, \ and\ \bibinfo {author}
  {\bibfnamefont {S.~W.}\ \bibnamefont {Hell}},\ }\href@noop {} {\bibfield
  {journal} {\bibinfo  {journal} {Proc. Nat. Acad. Sci. USA}\ }\textbf
  {\bibinfo {volume} {97}},\ \bibinfo {pages} {8206} (\bibinfo {year}
  {2000})}\BibitemShut {NoStop}%
\bibitem [{\citenamefont {Le~Sage}\ \emph {et~al.}(2013)\citenamefont
  {Le~Sage}, \citenamefont {Arai}, \citenamefont {Glenn}, \citenamefont
  {DeVience}, \citenamefont {Pham}, \citenamefont {Rahn-Lee}, \citenamefont
  {Lukin}, \citenamefont {Yacoby}, \citenamefont {Komeili},\ and\ \citenamefont
  {Walsworth}}]{DLS13}%
  \BibitemOpen
  \bibfield  {author} {\bibinfo {author} {\bibfnamefont {D.}~\bibnamefont
  {Le~Sage}}, \bibinfo {author} {\bibfnamefont {K.}~\bibnamefont {Arai}},
  \bibinfo {author} {\bibfnamefont {D.~R.}\ \bibnamefont {Glenn}}, \bibinfo
  {author} {\bibfnamefont {S.~J.}\ \bibnamefont {DeVience}}, \bibinfo {author}
  {\bibfnamefont {L.~M.}\ \bibnamefont {Pham}}, \bibinfo {author}
  {\bibfnamefont {L.}~\bibnamefont {Rahn-Lee}}, \bibinfo {author}
  {\bibfnamefont {M.~D.}\ \bibnamefont {Lukin}}, \bibinfo {author}
  {\bibfnamefont {A.}~\bibnamefont {Yacoby}}, \bibinfo {author} {\bibfnamefont
  {A.}~\bibnamefont {Komeili}}, \ and\ \bibinfo {author} {\bibfnamefont
  {R.~L.}\ \bibnamefont {Walsworth}},\ }\href@noop {} {\bibfield  {journal}
  {\bibinfo  {journal} {Nature}\ }\textbf {\bibinfo {volume} {496}},\ \bibinfo
  {pages} {486} (\bibinfo {year} {2013})}\BibitemShut {NoStop}%
\bibitem [{SM1()}]{SM1}%
  \BibitemOpen
  \href@noop {} {}\bibinfo {note} {See Supplemental Material for the derivation
  of the expression of the electric field and the conditional probability
  distribution.}\BibitemShut {Stop}%
\bibitem [{\citenamefont {Weitenberg}\ \emph
  {et~al.}(2011{\natexlab{b}})\citenamefont {Weitenberg}, \citenamefont
  {Schau\ss{}}, \citenamefont {Fukuhara}, \citenamefont {Cheneau},
  \citenamefont {Endres}, \citenamefont {Bloch},\ and\ \citenamefont
  {Kuhr}}]{WC112}%
  \BibitemOpen
  \bibfield  {author} {\bibinfo {author} {\bibfnamefont {C.}~\bibnamefont
  {Weitenberg}}, \bibinfo {author} {\bibfnamefont {P.}~\bibnamefont
  {Schau\ss{}}}, \bibinfo {author} {\bibfnamefont {T.}~\bibnamefont
  {Fukuhara}}, \bibinfo {author} {\bibfnamefont {M.}~\bibnamefont {Cheneau}},
  \bibinfo {author} {\bibfnamefont {M.}~\bibnamefont {Endres}}, \bibinfo
  {author} {\bibfnamefont {I.}~\bibnamefont {Bloch}}, \ and\ \bibinfo {author}
  {\bibfnamefont {S.}~\bibnamefont {Kuhr}},\ }\href {\doibase
  10.1103/PhysRevLett.106.215301} {\bibfield  {journal} {\bibinfo  {journal}
  {Phys. Rev. Lett.}\ }\textbf {\bibinfo {volume} {106}},\ \bibinfo {pages}
  {215301} (\bibinfo {year} {2011}{\natexlab{b}})}\BibitemShut {NoStop}%
\bibitem [{\citenamefont {Braginsky}\ and\ \citenamefont
  {Khalili}(1996)}]{BVB96}%
  \BibitemOpen
  \bibfield  {author} {\bibinfo {author} {\bibfnamefont {V.~B.}\ \bibnamefont
  {Braginsky}}\ and\ \bibinfo {author} {\bibfnamefont {F.~Y.}\ \bibnamefont
  {Khalili}},\ }\href {\doibase 10.1103/RevModPhys.68.1} {\bibfield  {journal}
  {\bibinfo  {journal} {Rev. Mod. Phys.}\ }\textbf {\bibinfo {volume} {68}},\
  \bibinfo {pages} {1} (\bibinfo {year} {1996})}\BibitemShut {NoStop}%
\bibitem [{SM2()}]{SM2}%
  \BibitemOpen
  \href@noop {} {}\bibinfo {note} {See Supplemental Material for some technical
  details.}\BibitemShut {Stop}%
\bibitem [{\citenamefont {Bauer}\ and\ \citenamefont {Bernard}(2011)}]{BM11}%
  \BibitemOpen
  \bibfield  {author} {\bibinfo {author} {\bibfnamefont {M.}~\bibnamefont
  {Bauer}}\ and\ \bibinfo {author} {\bibfnamefont {D.}~\bibnamefont
  {Bernard}},\ }\href {\doibase 10.1103/PhysRevA.84.044103} {\bibfield
  {journal} {\bibinfo  {journal} {Phys. Rev. A}\ }\textbf {\bibinfo {volume}
  {84}},\ \bibinfo {pages} {044103} (\bibinfo {year} {2011})}\BibitemShut
  {NoStop}%
\bibitem [{\citenamefont {Rau}\ \emph {et~al.}(2003)\citenamefont {Rau},
  \citenamefont {Dunningham},\ and\ \citenamefont {Burnett}}]{RAV03}%
  \BibitemOpen
  \bibfield  {author} {\bibinfo {author} {\bibfnamefont {A.~V.}\ \bibnamefont
  {Rau}}, \bibinfo {author} {\bibfnamefont {J.~A.}\ \bibnamefont {Dunningham}},
  \ and\ \bibinfo {author} {\bibfnamefont {K.}~\bibnamefont {Burnett}},\
  }\href@noop {} {\bibfield  {journal} {\bibinfo  {journal} {Science}\ }\textbf
  {\bibinfo {volume} {301}},\ \bibinfo {pages} {1081} (\bibinfo {year}
  {2003})}\BibitemShut {NoStop}%
\bibitem [{\citenamefont {Karski}\ \emph {et~al.}(2009)\citenamefont {Karski},
  \citenamefont {F\"orster}, \citenamefont {Choi}, \citenamefont {Alt},
  \citenamefont {Widera},\ and\ \citenamefont {Meschede}}]{KM09}%
  \BibitemOpen
  \bibfield  {author} {\bibinfo {author} {\bibfnamefont {M.}~\bibnamefont
  {Karski}}, \bibinfo {author} {\bibfnamefont {L.}~\bibnamefont {F\"orster}},
  \bibinfo {author} {\bibfnamefont {J.~M.}\ \bibnamefont {Choi}}, \bibinfo
  {author} {\bibfnamefont {W.}~\bibnamefont {Alt}}, \bibinfo {author}
  {\bibfnamefont {A.}~\bibnamefont {Widera}}, \ and\ \bibinfo {author}
  {\bibfnamefont {D.}~\bibnamefont {Meschede}},\ }\href {\doibase
  10.1103/PhysRevLett.102.053001} {\bibfield  {journal} {\bibinfo  {journal}
  {Phys. Rev. Lett.}\ }\textbf {\bibinfo {volume} {102}},\ \bibinfo {pages}
  {053001} (\bibinfo {year} {2009})}\BibitemShut {NoStop}%
\bibitem [{\citenamefont {Wenz}\ \emph {et~al.}(2013)\citenamefont {Wenz},
  \citenamefont {Zürn}, \citenamefont {Murmann}, \citenamefont {Brouzos},
  \citenamefont {Lompe},\ and\ \citenamefont {Jochim}}]{WAN13}%
  \BibitemOpen
  \bibfield  {author} {\bibinfo {author} {\bibfnamefont {A.~N.}\ \bibnamefont
  {Wenz}}, \bibinfo {author} {\bibfnamefont {G.}~\bibnamefont {Zürn}}, \bibinfo
  {author} {\bibfnamefont {S.}~\bibnamefont {Murmann}}, \bibinfo {author}
  {\bibfnamefont {I.}~\bibnamefont {Brouzos}}, \bibinfo {author} {\bibfnamefont
  {T.}~\bibnamefont {Lompe}}, \ and\ \bibinfo {author} {\bibfnamefont
  {S.}~\bibnamefont {Jochim}},\ }\href {\doibase 10.1126/science.1240516}
  {\bibfield  {journal} {\bibinfo  {journal} {Science}\ }\textbf {\bibinfo
  {volume} {342}},\ \bibinfo {pages} {457} (\bibinfo {year}
  {2013})}\BibitemShut {NoStop}%
\bibitem [{\citenamefont {Weiner}(2003)}]{JW03}%
  \BibitemOpen
  \bibfield  {author} {\bibinfo {author} {\bibfnamefont {J.}~\bibnamefont
  {Weiner}},\ }\href {http://dx.doi.org/10.1017/CBO9780511535215} {\emph
  {\bibinfo {title} {Cold and Ultracold Collisions in Quantum Microscopic and
  Mesoscopic Systems}}}\ (\bibinfo  {publisher} {Cambridge University Press},\
  \bibinfo {year} {2003})\BibitemShut {NoStop}%
\end{thebibliography}%

\widetext
\pagebreak
\begin{center}
\textbf{\large Supplemental Materials}
\end{center}

\renewcommand{\theequation}{S\arabic{equation}}
\renewcommand{\thefigure}{S\arabic{figure}}
\renewcommand{\bibnumfmt}[1]{[S#1]}
\renewcommand{\citenumfont}[1]{S#1}

\subsection{Expression of the electric field on the screen derived from the Heisenberg equation of motion}
The Hamiltonian under consideration is given by Eq. (1). Then, the  time evolution of the scattered field obeys the following Heisenberg equation of motion:  
\eqn{
\dot{\hat{a}}_{\mathbf{k'},\sigma}=-i\omega_{k'}\hat{a}_{\mathbf{k'},\sigma}+\frac{i}{\hbar}\sqrt{\frac{\hbar\omega_{k'}}{2\epsilon_{0}V}}\mathbf{e}_{\mathbf{k'},\sigma}\cdot\int d^{3}r'e^{-i\mathbf{k'\cdot\mathbf{r'}}}\Bigl[\mathbf{d}\cdot\hat{\Psi}_{g}^{\dagger}(\mathbf{r'})\hat{\Psi}_{e}(\mathbf{r'})+{\rm H.\,c.\, }\Bigr].
} 
Let $R_{\rm L}$ ($R_{\rm S}$) be the distance between the lens and the lattice (screen). We integrate out the Heisenberg equation of motion for the geometry shown in Fig. 1 and under the conditions $kR_{\rm L}, kR_{\rm S}\gg1$. Since the probe light is far-detuned, the excited state can be eliminated adiabatically, giving $\hat{\Psi}_{e}\simeq-\mathbf{d}^{*}\cdot E_{\rm P}^{(+)}\hat{\Psi}_{g}/\hbar\Delta$, where $\Delta$ is the detuning of the probe light. Then, after averaging out the polarization vector, we obtain the positive frequency component of the scattered field as follows:
\eqn{
\hat{E}_{\rm sca}^{(+)}(\mathbf{R})=\gamma\int d^{3}r'e^{-i\Delta\mathbf{k}\cdot\mathbf{r'}}\frac{J_{1}(ka\xi')}{ka\xi'}\hat{\Psi}_{g}^{\dagger}(\mathbf{r'})\hat{\Psi}_{g}(\mathbf{r'}),
}
where $a$ is the radius of the lens aperture, $\Delta\mathbf{k}$ is the wave-vector difference between incident and scattered photons, $\xi'$ is a geometrical factor defined by the angle spanned by the vectors $\mathbf{r'}$ and $\mathbf{R}$, and $\gamma$ is a coefficient determined by the angular momenta of atomic states and the polarization of the probe light. As a typical case, if we assume that the angular momenta of atomic states satisfy $m_{e}=m_{g}+1$ and the probe light is $\sigma^{+}-$polarized, the coefficient is given by
\eqn{
\gamma=\frac{-i\mathcal{E}_{0}k^{3}e^{ik(R_{\rm L}+R_{\rm S})}|d|^{2}}{8\pi\epsilon_{0}R_{\rm L}R_{\rm S}\hbar\Delta}\langle J_{e}m_{e}|J_{g}m_{g};11\rangle, 
}
where $\langle J_{e}m_{e}|J_{g}m_{g};11\rangle$ is the Clebsh-Gordan coefficient.
\subsubsection*{1D-case}
We can now derive the expression of the positive frequency component of the electric field on the screen, i.e., the measurement operator acting on the atomic state associated with photodetections on the screen. Let us first consider atoms trapped in a one-dimensional lattice and adopt the following tight-binding approximation: 
\eqn{
\hat{\Psi}_{g}(\mathbf{r})=\Phi(y,z)\sum_{m}w(x-md)\hat{b}_{m},
} 
where $\Phi(y,z)$ is the wave function confined in the transverse direction, $w(x)$ is the Wannier function centered at $x=0$. We assume that a length scale of our geometry is much larger than the lattice constant:  $R_{\rm L}, R_{\rm S}\gg{}d$. Under this assumption, the geometrical factor $\xi'$ takes a simple form, 
\eqn{
\xi'\simeq\frac{|x'-x_{X}|}{R_{\rm L}},
} 
where we introduce a diagonal position $x_{X}\equiv-R_{\rm L}X/R_{\rm S}$. After substituting these expressions into the scattered field and integrating the resultant expression with respect to $\mathbf{r'}$, we arrive at the expression of the operator associated with a photodetection at screen position $X$:
 \eqn{
 \hat{E}_{\rm sca}^{(+)}(X)=\gamma\sum_{m}e^{-i\Delta\mathbf{k}\cdot md\mathbf{e}_{x}}\mathcal{F}\Bigl[\frac{|x_{X}-md|}{\sigma}\Bigr]\hat{b}^{\dagger}_{m}\hat{b}_{m},
 }
which gives Eq. (2).
\subsubsection*{2D-case}
For a two-dimensional lattice, we can show, under $R_{\rm L}, R_{\rm S}\gg d$, that the geometrical factor $\xi'$ takes the form, 
\eqn{
\xi'\simeq\sqrt{\Bigl(\frac{|x'-x_{X}|}{R_{\rm L}}\Bigr)^{2}+\Bigl(\frac{|y'-y_{Y}|}{R_{\rm L}}\Bigr)^{2}},
}
where $y_{Y}\equiv-R_{\rm L}Y/R_{\rm S}$. Furthermore, we perform the tight-binding approximation and integrate the resultant scattered field with respect to $\mathbf{r'}$ in the same manner as in the 1D lattice. Then we obtain the following expression for the 2D lattice:
\eqn{
\hat{E}_{\rm sca}^{(+)}(X,Y)=\gamma\sum_{(m_{x},m_{y})}e^{-i\Delta\mathbf{k}\cdot\mathbf{m}}\mathcal{F}[\rho]\hat{b}_{m_{x},m_{y}}^{\dagger}\hat{b}_{m_{x},m_{y}}, 
}
where $\rho$ is defined by
\eqn{
\rho\equiv\sqrt{\Bigl(\frac{|x_{X}-m_{x}d|}{\sigma}\Bigr)^{2}+\Bigl(\frac{|y_{Y}-m_{y}d|}{\sigma}\Bigr)^{2}},
} 
and $(m_{x},m_{y})$ is a label of lattice sites, and $\mathbf{m}\equiv m_{x}d\mathbf{e}_{x}+m_{y}d\mathbf{e}_{y}$. This is the operator associated with a photodetection at position $(X,Y)$ on the screen. 
\subsection{Derivation of the conditional probability distribution}
We here mention the approximations made in deriving Eq. (4). We consider the probe light incident from a transverse direction. Then, the projected wave-vector difference between the incident and scattered photons in the direction of the lattice is by a factor of $\sim d/(R_{\rm L}+R_{\rm S})$ smaller than $k$ and the phase difference between light scattered from neighboring sites is negligibly small. Hence, the phase factor in Eq. (2) can be disregarded which results in a simple expression of the conditional probability distribution
\eqn{\label{psf}
P[X|\{n_{m}\}]=\frac{\bigl|\sum_{m=1}^{N_{\rm L}}n_{m}\mathcal{F}\bigl[\frac{|x_{X}-md|}{\sigma}\bigr]\bigr|^{2}}{\int dX'\bigl|\sum_{m=1}^{N_{\rm L}}n_{m}\mathcal{F}\bigl[\frac{|x_{X'}-md|}{\sigma}\bigr]\bigr|^{2}},
}
which gives Eq. (4).
\subsection{Derivation of the scaling behavior of the fidelity}
Here we conduct a scaling analysis of the fidelity and show the advantage of our method against an ordinary fitting procedure. To this end, we first consider the fidelity of our method. As mentioned in the main text, the fidelity is determined by the minimal value of the relative entropy between the photon-number distribution of different Fock states $R[\{n_{m}\}]$. Since we are here interested in the asymptotic scaling behavior, our problem of the scaling analysis of the fidelity reduces to the scaling analysis of the following slowest convergence rate $R_{\rm min}$:
\eqn{\label{mini}
R_{\rm min}\equiv\min_{\{n_{m}\}}R[\{n_{m}\}]=\min_{\{\{n_{m}\},\{n_{k}\}|\{n_{m}\}\neq\{n_{k}\}\}}D[P[X|\{n_{m}\}]|P[X|\{n_{k}\}]].
}  
Namely, $R_{\rm min}$ is the information-theoretic distance of the most indistinguishable Fock states. Below, we focus on the overlapped interference patterns which is of our primary interest.

 \begin{figure}
\scalebox{0.9}[0.9]{\includegraphics{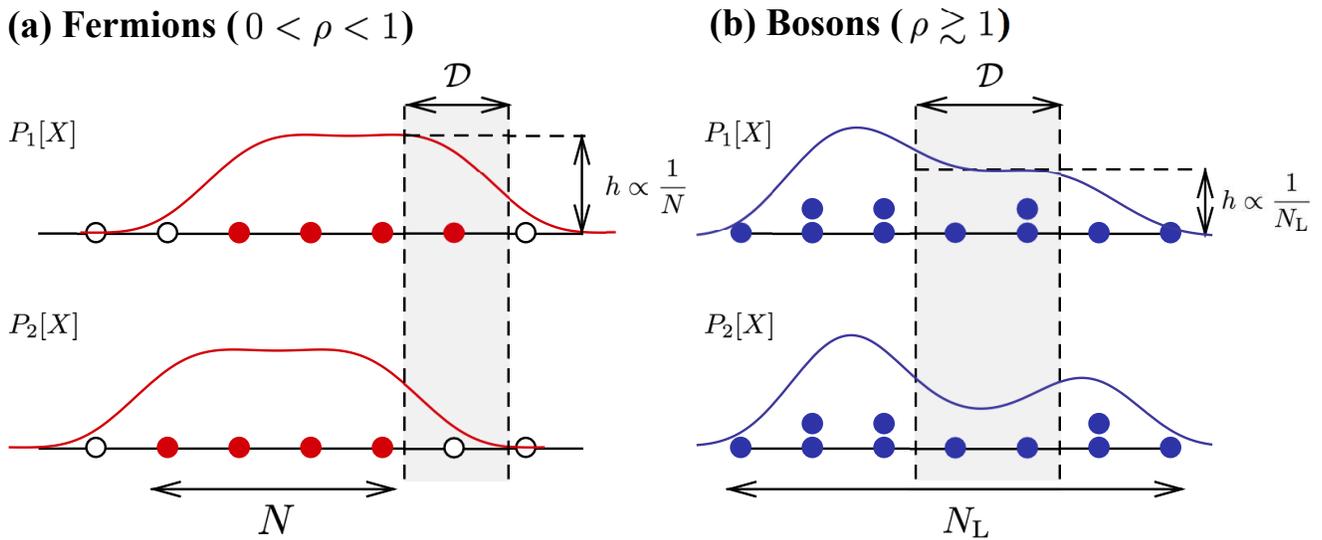}}
 \caption{\label{figS1}
Schematic illustrations of Fock states associated with the slowest convergence rate $R_{\rm min}$ for (a) fermions and (b) bosons. Here $P_{1,2}[X]$ denote the probability distributions of photodetections under consideration, $\mathcal{D}$ denotes the region contributing to the relative entropy between $P_{1,2}$, and $h$ is a typical scale of the height of the probability distribution in the region $\mathcal{D}$. $h$ scales with the inverse of the number of atoms $N$ (lattice sites $N_{\rm L}$) for (a) fermions ((b) bosons).
}
\end{figure}

\subsubsection*{Fermions}
Let us first consider fermions and simplify a discussion as follows. First of all, since the most indistinguishable Fock states of fermions are the closest-packed configurations in which only a single atom is misaligned (see Fig. \ref{figS1}), the dominant contribution to the relative entropy comes from a region in which (i) an atom is misaligned and the probability distributions are mismatched and (ii) the probability distribution under consideration takes a finite value. Then, we can approximate
\eqn{
R_{\rm min}=\int dX P_{1}[X]\ln\frac{P_{1}[X]}{P_{2}[X]}\propto\int_{\mathcal{D}}dX P_{1}[X]\sim h\cdot d,
}
where $P_{1,2}[X]$ are the photon-number distribution of the closest-packed Fock states in which a single atom is misaligned, $\mathcal{D}$ denotes a misaligned region and $h$ denotes a typical height of the probability distribution in the misaligned region $P_{1}[\mathcal{D}]$ (see Fig. \ref{figS1}). From the second to the third expression, we focus on the contributions from the misaligned region $\mathcal{D}$ and neglect a logarithmic dependence. In deriving the last expression, we approximate the value of the probability distribution $P_{1}$ in the misaligned region as constant $P_{1}\simeq h$ and the size of $\mathcal{D}$ as the lattice constant $d$. 

Due to these simplifications, the problem now reduces to the analysis of the scaling behavior of the height of the photon-number  distribution of the closest-packed Fock state. From the normalization condition of the probability distribution, we can easily obtain the following scaling relation $h\propto1/N$, where $N$ is the number of atoms. Hence, we conclude that $R_{\rm min}$ scales with the inverse of the atom number $R_{\rm min}\propto1/N$.
\subsubsection*{Bosons}
A similar discussion for bosons is possible when the filling factor $\rho\equiv N/N_{\rm L}$ is larger than one $\rho\gtrsim1$. Here $N_{\rm L}$ is the number of lattice sites. In this case, the most indistinguishable Fock states are the uniform configuration in which all lattice sites are occupied and an additional atom is misaligned (see Fig.\ref{figS1}). Roughly speaking, the height of the probability distribution $h$ is considered to scale with the number of lattice sites $N_{\rm L}$ from the normalization condition of the probability distribution and hence, along the lines of the discussion similar to fermions, we obtain the scaling relation $R_{\rm min}\propto1/N_{\rm L}$. For bosons with an intermediate filling fraction $0<\rho<1$, we find it difficult to conduct such a simple scaling analysis  because there is a numerous number of configurations for bosons and the problem of minimizing the relative entropy Eq.(\ref{mini}) seems to be highly complicated and non-trivial. 
\subsubsection*{Scaling behavior of our scheme}
To discuss the scaling behavior of our scheme, it is appropriate to consider the scaling of the error rate $\epsilon\equiv1-F$. We can show that the asymptotic behavior of $\epsilon$ is governed by the slowest exponentially decaying term and asymptotically given by $-\ln\epsilon\simeq N_{\rm p}R_{\rm min}$. Hence, together with the above discussions, we obtain the following scaling behavior of the error rate:
\eqn{\label{our}
\begin{cases}
-\ln\epsilon\propto \cfrac{N_{\rm p}}{N}&({\rm fermions}, 0<\rho<1),\\
\\
-\ln\epsilon\propto \cfrac{N_{\rm p}}{N_{\rm L}}&({\rm bosons}, \rho\gtrsim 1).
\end{cases}
}
We note that, for fermions, the exponential convergence rate of the error rate only depends on the number of detections per atom and hence, the achievable fidelity of our scheme is independent of details of configurations (e.g., the number of lattice sites and the filling factor). On the other hand, for bosons, the convergence rate of the error rate is determined by the number of detections per lattice site and depends on the configuration of the system considered.
\subsubsection*{A comparison with a fitting procedure}
To discuss the relation to an ordinary analysis, let us compare the scaling behavior of the ordinary fitting procedure with that of our model. To this end, by taking into account the discreteness of positions and the asymptotic normality of estimators, we model the fidelity of the fitting method as $F_{\rm fit}=({\rm erf}[d/(2\sqrt{2}\Delta_{n})])^{N}$ where $d$ is the lattice constant and $\Delta_{n}\equiv\Delta/\sqrt{n}$. Here $n\equiv N_{\rm p}/N$ is the number of signals per atom and $\Delta$ is a typical value of the width of the interference pattern. For example, if all atoms are sparsely distributed, then $\Delta\simeq\sigma$ and the above expression reduces to the joint probability of the event that each estimated position takes a value within the region corresponding to each lattice site compatible with the estimated Fock state.
Considering the overlapped interference patterns and the limit of a large number of photodetections, we can find the following scaling behavior of the error rate for the fitting procedure,
\eqn{\label{fit}
\begin{cases}
-\ln\epsilon_{\rm fit}\propto \cfrac{N_{\rm p}}{N}\cfrac{1-\rho}{1+\rho}&({\rm fermions},0<\rho<1),\\
\\
-\ln\epsilon_{\rm fit}\propto \cfrac{N_{\rm p}}{N_{\rm L}}\cfrac{1}{NN_{\rm L}}&({\rm bosons}, \rho\gtrsim 1).
\end{cases}.
}
By making a comparison between Eq. (\ref{our}) and Eq. (\ref{fit}), we conclude that our scheme achieves a faster convergence of the error rate than the fitting procedure for both fermions and bosons. Hence, for a given number of detections, the error rate of our method is exponentially smaller than that of the fitting method. In particular, we note that our scheme is especially advantageous in the high-density case in which the interference patterns significantly overlap and the ordinary fitting procedure almost fails to extract the position information as indicated by the above analysis.

Here one may expect that, since finding particles and finding holes seem to be identical tasks, advantageous situations of our method should be symmetric in the low and high fillings i.e., symmetric with respect to $\rho$ and $1-\rho$. While this should be the case for situations with a sufficiently high spatial resolution in which there exist no overlaps of the interference patterns from neighboring sites, we note that this is not true for situations with a low spatial resolution, which is of our interest. As discussed above and indicated by Eq. (\ref{fit}), in the low spatial resolution,  the regular fitting analysis almost fails to identify holes in the overlapped patterns because empty holes are significantly masked by signals coming from neighboring sites. However, our scheme based on Bayesian inference can efficiently extract information even if there exist overlaps and hence, our method remains valid in such situation as indicated by Eq. (\ref{our}). As a result, the advantageous situation of our method is not symmetric in the low and high filling regions for the low spatial resolution as considered in our analysis.  

\subsection{The least squares analysis}
We here explain details of the least squares analysis performed in our numerical simulations to show that the regular analysis also utilizes the knowledge about the discrete positions of atoms in a lattice. In the least squares analysis, we minimize the following $\chi$-square value with respect to all the Fock states $\{n_{m}\}$:
\eqn{
\chi^{2}(\{n_{m}\})=\sum_{i=1}^{N_{\rm pix}}\frac{(P[X_{i}|\{n_{m}\}]-P_{\rm meas}[X_{i}])^{2}}{\sigma_{i}^{2}}.
} 
Here $i$ denotes the label of each pixel, $N_{\rm pix}$ is the number of pixels, $X_{i}$ is the position of pixel $i$, and $P[X|\{n_{m}\}]$ denotes the probability distribution of photodetections when the atomic state is found to be the Fock state $\{n_{m}\}$. The measurement outcome is denoted by the renormalized histogram, $P_{\rm meas}[X_{i}]=N_{i}/N_{\rm photon}$, where $N_{i}$ is the number of photodetections at pixel $i$ and $N_{\rm photon}$ is the total number of photodetections. The weight in the denominator $\sigma_{i}^{2}$ is the variance of signals at pixel $i$ and is given by the expected signal value itself, $\sigma_{i}^{2}=P[X_{i}|\{n_{m}\}]$, because photodetection events are considered as a Poisson process. 

In the standard analysis conducted in our simulations, we calculate the above $\chi$-square value for all the Fock states and choose the most probable atomic configuration as the one that minimizes the value of the $\chi$-square. Here the crucial point is that, since the function $P[X|\{n_{m}\}]$ given by Eq. (\ref{psf}) already incorporates the knowledge of the discretized configuration of atoms in a lattice, the above regular analysis also utilizes the information of discrete atomic positions in the same manner as in our wavefunction collapse model.

\subsection{Reconstructing without prior knowledge of particle number}
We here mention the principle of why our scheme can be performed without prior knowledge about the total particle number. As discussed in the main text, we consider the relative entropy $D[P_{\rm meas}[X]|P[X|\{n_{m}\}]]$ between the measured distribution of photon-number $P_{\rm meas}[X]$ and the distribution corresponding to the collapsed Fock state $P[X|\{n_{m}\}]$. Since this information-theoretic measure converges to zero only if the assumed particle number is correct i.e., the measured distribution $P_{\rm meas}[X]$ approaches the true distribution $P[X|\{n_{m}\}]$, we can use this measure to determine the  total atom number. 

\begin{figure}
\scalebox{0.25}[0.25]{\includegraphics{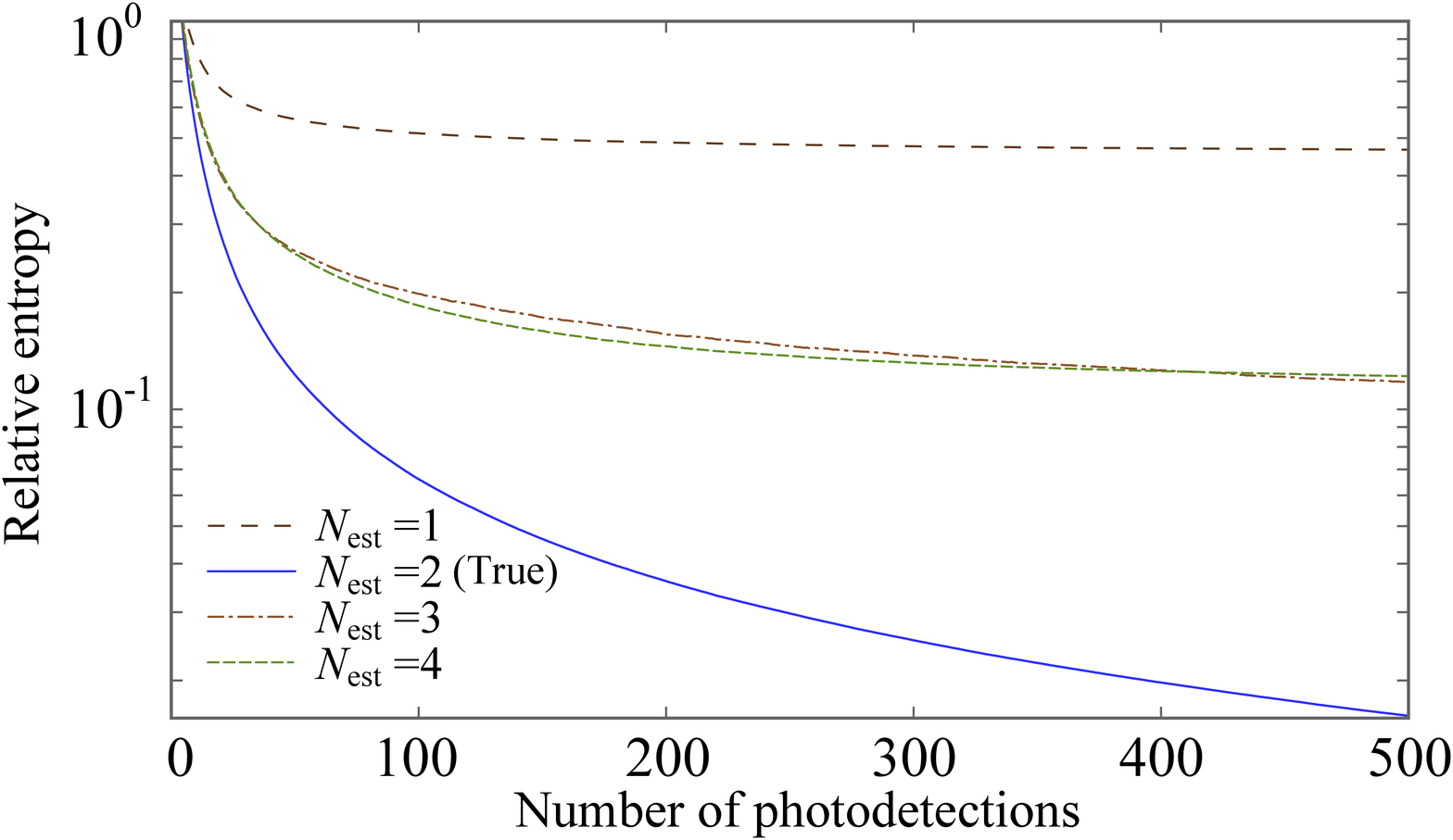}}
 \caption{\label{figS2}
Numerical results of the relative entropy between the measured distribution and the distribution corresponding to the collapsed Fock state for various assumed particle number $N_{\rm est}$. While the relative entropy converges to nonzero values for false particle numbers, it converges to zero for the true particle number as indicated by the blue solid curve. Here the average is taken over $10^4$ realizations for $\sigma=1.0$, $N_{\rm L}=5$ and $N=2$ fermions.
}
\end{figure}

To make quantitative discussions, we demonstrate this idea for two fermions by way of example as shown in Fig. \ref{figS2}. After only several tens of detections, the relative entropy associated with the true particle number (indicated by the blue solid curve) begins to apparently take the lower value relative to the other values which are based on the false particle numbers. While the relative entropy associated with the false particle number converges to some nonzero value, the value corresponding to the true particle number shows the clear convergence to zero. 

In practice, the number of detections is finite and a possible misestimation of the atom number may lead to a relatively small number of artificial deficits or surplus of atoms. For a more extensive analysis of this point and the generalization to the continuous positional space, see Ref. [52] in the main text.

\subsection{Recoil heating energy}
\begin{figure}
\scalebox{0.55}[0.55]{\includegraphics{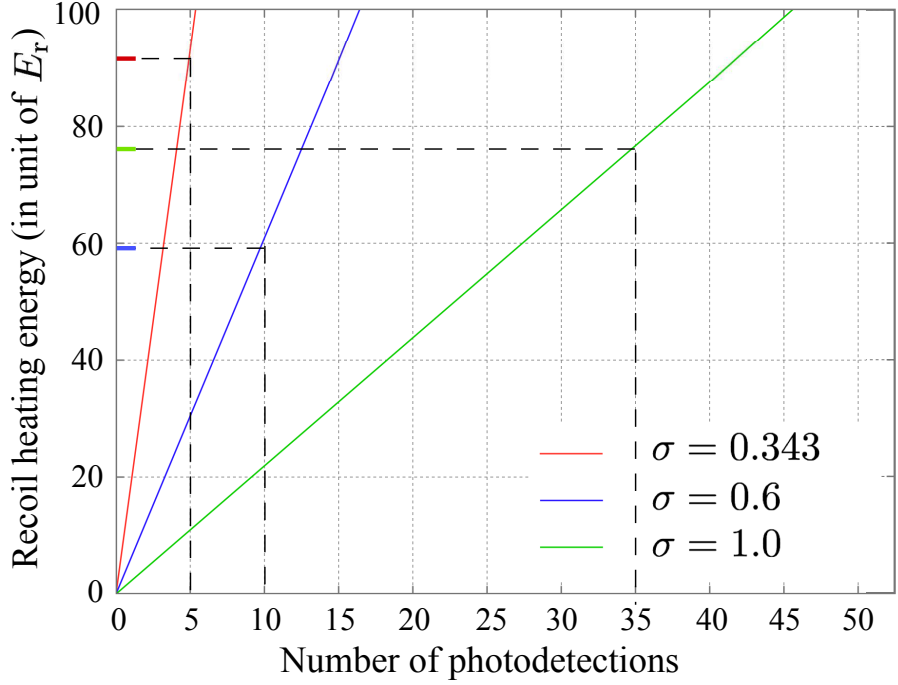}}
 \caption{\label{figS3}
 Average recoil heating energy per atom (in unit of $E_{r}$) associated with imaging light scattering against the number of photodetections for various resolution parameters $\sigma$ (in unit of $d$). The dashed lines indicate the required number of photodetections to achieve 95\% fidelity and the corresponding average heating energy.   
}
\end{figure}

Let us discuss an average recoil heating energy associated with  light scattering. The total heating energy per atom is estimated by the number of scattered photons and the recoil energy caused by imaging photons:
\eqn{
\frac{N_{\rm p}}{\eta}\cdot\frac{h^{2}}{2m\lambda^{2}}=\frac{N_{\rm p}}{\eta\pi^{2}N_{A}^{2}}\cdot\frac{d^{2}}{\sigma^{2}}\cdot E_{r},
}
where $N_{\rm p}$ is the number of detected photons, $\eta$ is the collection efficiency, $\lambda$ is the wavelength of the imaging light, $N_{A}$ is the numerical aperture, $d$ is the lattice constant, $\sigma=\lambda/2\pi N_{A}$ is the resolution parameter, and $E_{r}=\hbar^{2}\pi^{2}/2md^{2}$ is the recoil energy. For the sake of concreteness, we consider an experimental situation of Ref. [3] in the main text: $d=532$nm, $N_{A}=0.68$, and $\eta=0.1$. Figure \ref{figS3} shows the plot of the heating energy per atom (in unit of $E_{r}$) against the number of photodetections. The dashed lines indicate the required number of photodetections to achieve 95\% fidelity and the corresponding average heating energy.  While Ref. [3] uses the imaging light whose wavelength is $\lambda=780$nm (corresponding to $\sigma=0.343$ in unit of $d$), we also plot the results for longer imaging wavelengths corresponding to lower resolutions ($\sigma=0.6, 1.0$). In fact, for the lower-resolution cases, average heating energies (blue and green lines in Fig. \ref{figS3}) required to distinguish different Fock states are less than the average heating energy (red line in Fig. \ref{figS3}) for the current state-of-the-art resolution value. This is because the photon recoil energy decreases in proportional to the inverse square of  the imaging wavelength. We note that, in the very low resolution ($\sigma\geq1.0$, see Fig. 3(b) in the main text), the required number of photodetections rapidly increases and this increase eventually overwhelms the above advantage, and as a result, the average heating energy becomes large. Hence, there should be an optimal resolution where the average heating energy becomes minimal, and this optimal resolution value seems to be lower than the state-of-the-art resolution as indicated by Fig. \ref{fig3}.

\end{document}